\documentclass[useAMS,referee,usenatbib]{biom}
\newif\ifbiomreferee
\IfClassLoadedWithOptionsTF{biom}{referee}
  {\biomrefereetrue}
  {\biomrefereefalse}

\usepackage{amsmath,amsfonts,amssymb}
\DeclareMathAlphabet{\mathbbn}{U}{bbold}{m}{n}

\usepackage[dvips]{graphicx}
\usepackage[dvipsnames,table]{xcolor}
\usepackage{array}
\usepackage{pdfpages}   % used to concat the supplement pdf and main pdf
\usepackage{collcell}   % for certain table formatting tasks
\usepackage{xr-hyper}
\usepackage{hyperref}   % for intra-PDF hyperlinks 
    \hypersetup{ 
      colorlinks   = true, %Colours links instead of ugly boxes
      urlcolor     = black, %Colour for external hyperlinks
      linkcolor    = black, %Colour of internal links
      citecolor   = black %Colour of citations
    }
\usepackage[nameinlink,noabbrev]{cleveref}
    \crefname{webappendix}{appendix}{appendices}
        \Crefname{webappendix}{Appendix}{Appendices}
        \crefformat{webappendix}{#2appendix~#1#3}
        \Crefformat{webappendix}{#2Appendix~#1#3}
    \crefname{model}{model}{models}
        \Crefname{model}{Model}{Models}
        \crefformat{model}{model~#2(#1)#3} % Produces eq. (1) with (1) as a link
        \Crefformat{model}{Model~#2(#1)#3} % Produces eq. (1) with (1) as a link
        
    \crefrangelabelformat{section}{#3#1#4--#5\crefstripprefix{#1}{#2}#6}

\newcommand{\func}[2]   { \ensuremath\mathrm{#1}\left(#2\right) }

\renewcommand{\Pr}[1]   { \func{Pr}{#1}     }

\newcommand{\E}[2]      { \ensuremath\mathbb{E}_{#1}\left(#2\right) }

\newcommand{\Var}[1]    { \func{Var}{#1}    }

\newcommand{\Cor}[1]    { \func{Cor}{#1}    }

\renewcommand{\log}[1]  { \func{log}{#1}            }

\newcommand{\Bin}[1]        { \func{Binom}{#1}          }

\newcommand{\Normal}[1]     { \func{N}{#1}              }

\renewcommand{\Gamma}[1]    { \func{Gamma}{#1}          }

\newcommand{\iid}       { \overset{\small\mathrm{iid}}{\sim}    }
\newcommand{\indepsim}  { \overset{\smash{\scriptstyle\mathrm{indep}}}{\sim}}

\let\bm\bmath % also italicizes bold characters. only works with AMSMath, I think

\DeclareUnicodeCharacter{0394}{\ensuremath{\Delta}}
\DeclareUnicodeCharacter{0398}{\ensuremath{\Theta}}
\DeclareUnicodeCharacter{039B}{\ensuremath{\Lambda}}
\DeclareUnicodeCharacter{03A0}{\ensuremath{\Pi}}
\DeclareUnicodeCharacter{03A3}{\ensuremath{\Sigma}}
\DeclareUnicodeCharacter{03A6}{\ensuremath{\Phi}}
\DeclareUnicodeCharacter{03A8}{\ensuremath{\Psi}}
\DeclareUnicodeCharacter{03A9}{\ensuremath{\Omega}}

\DeclareUnicodeCharacter{03B1}{\ensuremath{\alpha}}
\DeclareUnicodeCharacter{03B2}{\ensuremath{\beta}}
\DeclareUnicodeCharacter{03B3}{\ensuremath{\gamma}}
\DeclareUnicodeCharacter{03B4}{\ensuremath{\delta}}
\DeclareUnicodeCharacter{03B7}{\ensuremath{\eta}}
\DeclareUnicodeCharacter{03B8}{\ensuremath{\theta}}
\DeclareUnicodeCharacter{03BA}{\ensuremath{\kappa}}
\DeclareUnicodeCharacter{03BB}{\ensuremath{\lambda}}
\DeclareUnicodeCharacter{03BC}{\ensuremath{\mu}}
\DeclareUnicodeCharacter{03C0}{\ensuremath{\pi}}
\DeclareUnicodeCharacter{03C1}{\ensuremath{\rho}}
\DeclareUnicodeCharacter{03C3}{\ensuremath{\sigma}}
\DeclareUnicodeCharacter{03C4}{\ensuremath{\tau}}
\DeclareUnicodeCharacter{03C6}{\ensuremath{\phi}}
\DeclareUnicodeCharacter{03D5}{\ensuremath{\phi}}
\DeclareUnicodeCharacter{03C8}{\ensuremath{\psi}}
\DeclareUnicodeCharacter{03C9}{\ensuremath{\omega}}

\DeclareUnicodeCharacter{0D7}{\ensuremath{\times}}
\DeclareUnicodeCharacter{02208}{\ensuremath{\in}}
\DeclareUnicodeCharacter{0221D}{\ensuremath{\propto}}
\DeclareUnicodeCharacter{02248}{\ensuremath{\approx}}
\DeclareUnicodeCharacter{0221E}{\ensuremath{\infty}}
\DeclareUnicodeCharacter{02261}{\ensuremath{\equiv}}
\DeclareUnicodeCharacter{02264}{\ensuremath{\leq}}
\DeclareUnicodeCharacter{02265}{\ensuremath{\geq}}
\DeclareUnicodeCharacter{02115}{\ensuremath{\mathbb{N}}}
\DeclareUnicodeCharacter{0211D}{\ensuremath{\mathbb{R}}}

\makeatletter
\newcommand{\smallbullet}{} % for safety
\DeclareRobustCommand\smallbullet{%
  \mathord{\mathpalette\smallbullet@{0.6}}%
}
\newcommand{\smallbullet@}[2]{%
  \vcenter{\hbox{\scalebox{#2}{$\m@th#1\bullet$}}}%
}
\makeatother
\DeclareUnicodeCharacter{02219}{\ensuremath{\smallbullet}}

\AtBeginDocument{%
  \setlength{\hoffset}{0pt}%
}

\definecolor{mylightgray}{RGB}{230,230,230}

\title[Mixing and Heterogeneity in Surveillance Models]{Bayesian Inference of Mixing and Transmission Heterogeneity in Stratified Disease Surveillance Models}

\author{Miles Moran$^*$\email{miles.moran@oregonstate.edu}, 
Rob Trangucci$^{**}$\email{rob.trangucci@oregonstate.edu}, and 
Lisa Madsen$^{***}$\email{lisa.madsen@oregonstate.edu} \\
Department of Statistics, Oregon State University, Corvallis, Oregon, USA}

\begin{document}

% \date{{\it Received [month]} [year]. {\it Revised [month]} [year].  {\it
% Accepted [month]} [year].}

\pagerange{\pageref{firstpage}--\pageref{lastpage}} 
\volume{[volume]}
\pubyear{[year]}
\artmonth{[month]}
\doi{???/???.???}

\label{firstpage}

\begin{abstract}
    When surveillance data of infectious disease incidence (e.g. weekly case counts) are disaggregated by demographic indicators, disparities in long-run health outcomes between these groups become apparent. Accurate identification of high-risk subpopulations would enable policy-makers to target interventions early in an epidemic; but, temporal models of disease incidence typically lack robust treatment of multivariate (i.e. subpopulation-level) outcomes. We propose a novel Bayesian latent-variable extension of the \textit{endemic-epidemic} (``EE'') modeling framework commonly used for this purpose. Specifically, we augment the EE model class with explicit representation of unobserved individual-level transmissibility; explicit separation of disease incidence and prevalence; and parametric estimation of between-demographic-groups mixing structure. The resulting model may be tailored for either rare-disease (highly-endemic) contexts or outbreak-driven (highly-epidemic) contexts, and is capable of inferring social contact mixing patterns from incidence data alone, including mixing patterns among multiply-stratified data. To demonstrate, we conduct a simulation study comparing our model to an existing doubly-stratified EE model in the intended rare-disease application regime. We then compare our inference to the competitor’s for real incidence data of norovirus gastroenteritis in Berlin, 2011-2015, disaggregated by six age groups and twelve geographic regions. Finally, we report inference of our model on COVID-19 incidence recorded in Michigan during the first year of the pandemic, disaggregated by six age groups and sixty-six geographic regions. 
    \vspace{10mm}
\end{abstract}


\maketitle

%%%%%%%%%%%%%%%%%%%%%%%%%%%%%%%%%%%%%%%%%%%%%%%%%%%%%%%%%%%%%%%%%%%%%%%%%%%%%%%%
%%%%%%%%%%%%%%%%%%%%%%%%%%%%%%%%%%%%%%%%%%%%%%%%%%%%%%%%%%%%%%%%%%%%%%%%%%%%%%%%

%%%%%%%%%%%%%%%%%%%%%%%%%%%%%%%%%%%%%%%%%%%%%%%%%%%%%%%%%%%%%%%%%%%%%%%%%%%%%%%%
%%%%%%%%%%%%%%%%%%%%%%%%%%%%%%%%%%%%%%%%%%%%%%%%%%%%%%%%%%%%%%%%%%%%%%%%%%%%%%%%

\section{Introduction}
\label{s:intro}

Modern disease surveillance systems record incident counts stratified by multiple factors, typically geography combined with a demographic such as age. These subpopulation-level data enable investigation of disease transmission and mixing dynamics; however, fine partitioning produces sparse, unreplicated time series. The resulting marginal distributions for subpopulation incidence are characterized by extreme zero-inflation and occasional large events. These patterns are poorly captured by large-population approximations (e.g. deterministic compartment models with Gaussian errors) but may be consistent with models of incidence featuring individual-level variation in infectiousness \citep{2005-Lloyd-Smith-Superspreading}. We use the phrase ``transmission heterogeneity'' to refer to this theoretical basis. At the same time, each subpopulation contributes at most one observation per time step, while the number of subpopulations to model grows multiplicatively in the number of stratifications. In these settings, model flexibility is constrained by weak identifiability, and fitting strategies often rely on rigid structural assumptions \citep{2017-Meyer-Held-Social-Contact,2018-Bauer-Wakefield-ID-Model-HFMD}. 

Adequate inference of subpopulation-level transmission and mixing patterns requires a model capable of addressing these issues. Thoughtful subpopulation-level modeling may also benefit analyses where between-group dynamics are not of primary concern by enabling researchers to define population-level estimands that are robust to ecological bias. Recent work has quantified such bias for next-generation-matrix estimation of $R_0$ (a population's basic reproduction number) in deterministic compartment models \citep{2024-Manna-Generalized-Contact}. Unfortunately, analogous examination of discrete-time stochastic models of surveillance data remain limited, particularly in doubly stratified settings. All this motivates exploration of subpopulation-level models that (i) bear parameters interpretable on the scale of incident counts, (ii) allow mixing parameters to be learned from incidence data alone, and (iii) remain usable when subpopulation sizes are small enough for individual-level effects to surface. 

Existing discrete-time stochastic models of disease incidence often align with one of three model classes. These include time series susceptible-infected-recovered (``TSIR'') models and their related renewal-theory formulations \citep{2000-Finkenstadt-Grenfell-TSIR, 2002-Bjornstad-Measles-TSIR}, endemic-epidemic (``EE'' or \texttt{hhh4}) models \citep{2005-Held-Surveillance-Counts, 2012-Held-Paul-Modeling-Seasonality}, and binomial chain models \citep{1999-Daley-Gani-Epidemic-Modelling}. Each framework has seen extensive application in the context of univariate data and surveillance data singly-stratified by geographic unit \citep{2017-Meyer-Held-Hohle-Surveillance-Package, 2019-Bjornstad-Grenfell-Human-Movement}, including in Bayesian inference paradigms \citep{2003-KnorrHeld-Richardson-Bayesian-Endemic, 2023-Jewell-Bayesian-Binomial-Chain, 2025-Adeoye-Bayesian-Outbreak-Detection}. 

The TSIR and EE frameworks are attractive for modeling surveillance data because they yield interpretable mean structures and can be estimated with standard likelihood-based tools; however, they are not universally applicable. These models may be poorly calibrated when incidence is high in small subpopulations, due to the lack of upper-bounding on the mean parameter and on the support of the response. This alone makes the TSIR and EE models tenuous choices for most doubly-stratified surveillance data. Binomial chain models naturally address this limitation but are analytically complex and lose interpretation of model parameters on the scale of incident counts. Still, any extension of these models to doubly-stratified data would require bespoke methods for estimation of subpopulation mixing parameters. Moreover, all three model classes typically assume independence between subpopulation counts after conditioning on lag-1 incidence. When individual-level transmission is heterogeneous, e.g., when a small fraction of infectious individuals generate a disproportionate number of secondary cases, the resulting time series may feature clustering and heavy tails not well-represented by observation-level overdispersion alone. 

Recent models explicitly represent transmission heterogeneity \citep{2022-Zhang-Transmission-Heterogeneity-Surveillance, 2025-Craddock-Bayesian-Super-Spreading} in response to the seminal work of Lloyd-Smith and company. A singly-stratified variant is presented in \citet{2020-Zelner-Transmission-Heterogeneity} for description of super-spreading-aware spatial transmission, but no such analog exists for doubly-stratified data. Our contribution, then, can be summarized as an introduction of these new models to doubly-stratified settings.

We propose a distribution-agnostic, latent variable generalization of the endemic-epidemic class of models, motivated by the investigation of transmission heterogeneity in \citet{2005-Lloyd-Smith-Superspreading} and \citet{2020-Zelner-Transmission-Heterogeneity} and intended for use with doubly-stratified surveillance data. Like the EE model framework, our proposed model links mean incidence to a linear predictor with ``endemic'' and ``epidemic'' components; however, our proposed linear predictor explicitly represents realized generation-level infectiousness, rather than a deterministic function of lagged incidence. This construction nests standard EE models as a limiting case, but importantly induces cross-subpopulation covariance that is otherwise missing. We fit the proposed model with Hamiltonian Monte Carlo (HMC) using Stan \citep{2025-Stan} and \texttt{CmdStanR} \citep{2025-CmdStanR}. Fitting the model in a Bayesian paradigm enables estimation of mixing patterns with looser parametric assumptions than existing methods and avoids the computation of marginalizing over the latent layer. 

The remainder of this paper is organized as follows. \Cref{s:model} introduces the proposed doubly-stratified latent-infectiousness model. We present the model in its most general form and use key marginal moment expressions to clarify the nesting relationship of our model to standard EE models. \Cref{s:simstudy} begins with a specific model instance tailored to rare-disease contexts then details a simulation study assessing its forecasting and inference capabilities against an EE model comparator. \Cref{s:inf-noro} sees application of the proposed rare-disease model instance to incident counts of Norovirus Gastroenteritis and compares our inference to that of \citet{2017-Meyer-Held-Social-Contact}. \Cref{s:inf-covid} begins with a specific instance of the proposed model tailored to outbreak-driven epidemics in finely-partitioned datasets, then applies this model instance to a novel dataset of early-pandemic COVID-19 incidence in Michigan. We conclude with a discussion of our findings in \Cref{s:discuss}. 

%%%%%%%%%%%%%%%%%%%%%%%%%%%%%%%%%%%%%%%%%%%%%%%%%%%%%%%%%%%%%%%%%%%%%%%%%%%%%%%%
%%%%%%%%%%%%%%%%%%%%%%%%%%%%%%%%%%%%%%%%%%%%%%%%%%%%%%%%%%%%%%%%%%%%%%%%%%%%%%%%
%%%%%%%%%%%%%%%%%%%%%%%%%%%%%%%%%%%%%%%%%%%%%%%%%%%%%%%%%%%%%%%%%%%%%%%%%%%%%%%%
%%%%%%%%%%%%%%%%%%%%%%%%%%%%%%%%%%%%%%%%%%%%%%%%%%%%%%%%%%%%%%%%%%%%%%%%%%%%%%%%

\section{Model}
\label{s:model}

\subsection{Latent Infectiousness Framework}
\label{ss:model-def-general}

Index observation time (generation) by $t∈\{1,...,T\}$, geographic region by $g∈\{1,...,G\}$, and age group by $i∈\{1,...,I\}$. For cell $(g,i)$, let $(X_{tgi},Y_{tgi})$ denote the true number of susceptible and infectious individuals at time $t$, and let $Y^{*}_{tgi}$ denote the number of cases newly-reported during the period $(t,t+1)$. We assume perfect ascertainment and thus use ``incidence'' and ``\textit{reported} incidence'' interchangeably hereafter.

We assume each infectious individual $j$ from cell $(g,i)$ generates a latent infectious potential \mbox{$r_{tgij}\iid\mathrm{Gamma}(\tfrac{R_{tgi}}{θ},θ)$} during the period $(t,t+1)$, with $\mathrm{Gamma}(\cdot,\cdot)$ given in shape-scale notation. The resulting subpopulation aggregate \mbox{$r_{tgi} = \sum_{j}^{Y_{tgi}} r_{tgij}$} is again Gamma distributed (or a known zero when $Y_{tgi}=0$). This provides a subpopulation-level latent variable 
\begin{equation}
    (r_{tgi} \mid Y_{tgi})
        \indepsim \mathrm{Gamma}\left( \frac{R_{tgi}Y_{tgi}}{θ}, θ \right)
\label{eq:model-general-1}
\end{equation}
which may serve as a transmission-heterogeneity-aware substitute for infectious prevalence $Y_{tgi}$ in a formulation of mean incidence. To this end, we describe the allocation of infectiousness contributions from all source cells $\{(g',i')\}$ specifically to recipient cell $(g,i)$ through a nonnegative linear predictor $λ_{tgi}$. Using $\bm{r}_{t}$ to denote the $G×I$ matrix with entries $r_{tgi}$, write
\begin{equation}
    λ_{tgi}(\bm{r}_{t})
        = δ_{tgi} + ϕ_{tgi}\sum_{g',i'} w^{(G)}_{gg'} w^{(I)}_{ii'} r_{tg'i'} \, . 
\label{eq:model-general-2}
\end{equation}
As in the EE model framework, the ``endemic'' term $δ_{tgi}>0$ reflects infectiousness contributions imported from outside the study population, and the ``epidemic'' term reflects the interaction between a subpopulation's natural susceptibility $ϕ_{tgi}>0$ and its incoming intra-population contributions $\bm{r}_t$. We use \mbox{$w^{(G)}_{gg'}w^{(I)}_{ii'}$} to denote the proportion of expected secondary cases generated by subpopulation $(g',i')$ that is sent to subpopulation $(g,i)$. These are weights over recipients, i.e. \mbox{$\sum_{g} w^{(G)}_{gg'}=\sum_{i} w^{(I)}_{ii'}=1$}. 

The latent infectious potential (or just ``infectiousness'') variables $r_{tgij}$ and their cell-level aggregates are similar to the ``instant individual reproduction numbers'' described in \citet{2022-Zhang-Transmission-Heterogeneity-Surveillance}, but $r_{tgij}$ need not bear interpretation as an individual's expected secondary cases except in those specific models where mean incidence is linear in the $r_{tgi}$ contributions. We consider a broader class of models here and thus withhold the specific interpretation of $r_{tgi}$ and $λ_{tgi}$ until a concrete model instance is defined. 

What remains to instantiate a model is a conditional likelihood \mbox{$F(Y^{*}_{tgi}|X_{tgi},\bm{r}_{t})$} and ``link'' function $g$ tying mean incidence to infectiousness contributions and susceptible pool size. We require that mean incidence increase monotonically as a function of $λ_{tgi}$ and write
\begin{equation}%
    \E{}{Y^{*}_{tgi} \mid X_{tgi}, \bm{r}_{t}}
        = g(X_{tgi}, λ_{tgi}(\bm{r}_t))
\label{eq:model-general-3}
\end{equation}
but do not advise a specific class of $g$ for all model instances. Adjustment of the mean for small subpopulation sizes $X_{tgi}$ is salient in doubly-stratified surveillance data. In this case, we default to smooth $g$ satisfying $\lim_{λ_{tgi}\uparrow\infty}g(X_{tgi},λ_{tgi})=X_{tgi}$ and $g(0,λ_{tgi}) = g(X_{tgi},0) = 0$ in addition to the previous monotonicity condition, and we combine $g$ with a conditional likelihood of bounded support (Binomial, Quasi-Binomial, etc.). These choices of $g$ and $F$ complete the model when when the true number of susceptible and infectious individuals, $X_{tgi}$ and $Y_{tgi}$, are known. As this is rarely the case for surveillance data, we use plug in estimates $\hat{X}_{tgi}$ and $\hat{Y}_{tgi}$ computed from data according to some bespoke function $h_{\bm{γ}}$ (which may involve unknown parameters $\bm{γ}$). In the following \Cref{ss:model-def-instances}, we motivate the two model instances presented in this manuscript as special cases of the general model form. 

%------------------------------------------------------------------------------%

\subsection{Model Instances Derived from General Form}
\label{ss:model-def-instances}

Natural choices for a bounded link $g$ and conditional likelihood $F$ arise in the EE model formulation presented in \citet[eq.~23.7]{2019-Wakefield-Book-Surveillance-Data} and the renewal process formulation presented in \citet[eq.~6]{2023-Bhatt-Renewal-Processes}. If we follow these, then $λ_{tgi}$ enters the model as an interval-level cumulative infection hazard (i.e. a within-interval cumulative \textit{force of infection}, ``FOI'') in the binomial chain likelihood \mbox{$(Y^{*}_{tgi}|X_{tgi},\bm{r}_{t})\sim \mathrm{Binom}(X_{tgi}, 1-e^{-λ_{tgi}(\bm{r}_t)})$}. The choices \mbox{$g(X_{tgi},λ_{tgi})=X_{tgi}(1-e^{-λ_{tgi}})$} and \mbox{$F\sim\mathrm{Binom}(\cdot,\cdot)$} adjust the linear predictor $λ_{tgi}(\bm{r}_t)$ to a finite population size \textit{and} bestow it with an epidemiologically-grounded interpretation. This is, in essence, the ``outbreak-driven'' model we fit to \mbox{COVID-19} data in \Cref{s:inf-covid} after introducing assumptions of permanent susceptible depletion \mbox{$\hat{X}_{tgi} = E_{gi} - \sum_{d=1}^{t-1} Y^{*}_{(t-d)gi}$} and prevalence from lasting infection \mbox{$\hat{Y}_{tgi} = \sum_{d=1}^{t-1} e^{-γ(d-1)}Y^{*}_{(t-d)gi}$}.

Alternatively, we may choose $g$ and $F$ in closer accordance with the EE model class presented in \citet{2017-Meyer-Held-Hohle-Surveillance-Package}, wherein $λ_{tgi}$ enters the model as a direct approximation of mean incidence, \mbox{$(Y^{*}_{tgi}|\bm{r}_{t})\sim \mathrm{NegBin}(λ_{tgi}(\bm{r}_t), ψ)$}, with $ψ$ satisfying \mbox{$\Var{Y^{*}_{tgi}|\bm{r}_{t}} = λ_{tgi}(1+ψλ_{tgi})$}. The choices \mbox{$g(X_{tgi},λ_{tgi})=λ_{tgi}$} and \mbox{$F\sim\mathrm{NegBin}(\cdot,\cdot)$} leave the linear predictor unadjusted, reducing the analytical complexity of the model and restoring the interpretation of $r_{tgij}$ as individual-level expected secondary cases. Broadly, this is a transmission-heterogeneity version of the classic EE model for rare-disease settings. We explore this relationship in \Cref{ss:rel-to-ee} and fit such a model in \Cref{s:simstudy,s:inf-noro}.

We introduce the proposed model in its most general form to emphasize its flexibility and relationship to other frameworks. Lesser-explored choices of $g$ and $F$ may also arise naturally, e.g. use of \mbox{$g(X_{tgi},λ_{tgi})=X_{tgi}(1-e^{-λ_{tgi}^{k}})$} in outbreak-driven models with Weibull-distributed times to infection. Importantly, when the dispersion parameter $θ$ of the assumed latent variable distribution is taken $θ\downarrow 0$, the $r_{tgi}$ form a point-mass about their means $R_{tgi}Y_{tgi}$, thus creating a natural null model for identifying the presence of transmission heterogeneity. 

%%%%%%%%%%%%%%%%%%%%%%%%%%%%%%%%%%%%%%%%%%%%%%%%%%%%%%%%%%%%%%%%%%%%%%%%%%%%%%%%
%%%%%%%%%%%%%%%%%%%%%%%%%%%%%%%%%%%%%%%%%%%%%%%%%%%%%%%%%%%%%%%%%%%%%%%%%%%%%%%%

\subsection{Nesting Relationship to Endemic-Epidemic Models in Rare Disease Contexts}
\label{ss:rel-to-ee}

We now clarify how the proposed model generalizes the EE framework, and we contrast their resulting marginal moment structures. In \Cref{ss:model-def-general}, we framed the link $g$, likelihood $F$, and population estimation function $h_{\bm{γ}}$ as modeling ``choices.'' In a rare disease context, however, the natural choices for $g$, $F$, and $h_{\bm{γ}}$ can be re-framed as an analytical reduction of our proposed model to the classical EE model under select simplifying assumptions. 

For a disease with low incidence per capita, we may first assume the depletion of the susceptible pool is negligible, \mbox{$\hat{X}_{tgi} = E_{gi}$}. Next, assume the infectious period is brief enough to estimate prevalence equal to incidence, \mbox{$\hat{Y}_{tgi} = Y^{*}_{(t-1)gi}$}. Assuming transmission heterogeneity effects are negligible ($θ≈0$), the infectious load $λ_{tgi}$ becomes a deterministic function of \mbox{lag-1} incidence. Any link $g$ should map zero infectious load to zero expected incidence, increase in $\lambda_{tgi}$, and be locally identity on the infectious-load scale. Thus, for sufficiently rare incidence, \mbox{$g(\hat{X}_{tgi}, λ_{tgi})=λ_{tgi} + o(λ_{tgi}) ≈ λ_{tgi}$}, making the finite-population adjustment negligible.
Since \mbox{$\E{}{Y^{*}_{tgi}|λ_{tgi}} = λ_{tgi}$} is unbounded, we must choose a likelihood with support $ℕ$, e.g. negative binomial. Substituting these pieces into \Crefrange{eq:model-general-1}{eq:model-general-3} yields
\begin{equation*}
    \E{θ=0}{Y^{*}_{tgi} \mid \bm{Y}^{*}_{t-1}}
        = δ_{tgi} + R_{0} ϕ_{tgi}\sum_{g',i'} w^{(G)}_{gg'} w^{(I)}_{ii'} Y^{*}_{(t-1)g'i'}.
\end{equation*}
Up to notation, this is exactly the doubly-stratified variant of the EE model presented in \citet[eq.~10]{2017-Meyer-Held-Hohle-Surveillance-Package}.

Consider this EE model alongside its immediate extension where $θ>0$ (still retaining all other rare-disease assumptions). The introduction of latent transmission heterogeneity does not impact marginal mean incidence, as $λ_{tgi}(\bm{r}_t)$ is linear in $\bm{r}_{t}$ (see \Cref{WA:proof-means}).
The law of total variance and conditional independence of $r_{tgi}$ yields
\begin{equation*}
    \mathrm{Var}_{θ>0}\left(Y^{*}_{tgi}\mid \bm{Y}^{*}_{t-1}\right)
        = \mathrm{Var}_{θ=0}\left(Y^{*}_{tgi}\mid \bm{Y}^{*}_{t-1}\right) + θ \mathbb{V} + θψ\mathbb{V}, 
\end{equation*}
where $\mathbb{V} = R_0 ϕ_{tgi}^2 \sum_{g',i'} \{w^{(G)}_{gg'} w^{(I)}_{ii'}\}^{2} Y^{*}_{(t-1)g'i'}$ is invariant to $θ$ and $ψ$. In other words, the excess marginal variance induced by way of transmission heterogeneity scales non-linearly as the product of $θ$ and $ψ$. 

Similarly, in our framework, marginalizing over the latent $r_{tgi}$ induces conditional dependence across recipient subpopulations, in contrast to the EE framework, which assumes independence conditional on lag-1 incidence.
Formally, for two recipient cells $(g,i)$ and $(a,b)$, we have
\ifbiomreferee
    \begin{equation*}
        \mathrm{Cov}(Y^*_{tgi}, Y^*_{tab}|\bm{Y}^{*}_{t-1})
            = θ R_0 ϕ_{tgi} ϕ_{tab}
                \sum_{g',i'}  
                    w^{(G)}_{gg'} w^{(I)}_{ii'} w^{(G)}_{ag'} w^{(I)}_{bi'}
                    Y^{*}_{(t-1)g'i'}.
    \end{equation*}
\else 
    \begin{equation*}
        \mathrm{Cov}(Y^*_{tgi}, \! Y^*_{tab}|\bm{Y}^{*}_{t-1})
            \! = \! θ R_0 ϕ_{tgi}ϕ_{tab} \!
                \sum_{g',i'} \! 
                    w^{(G)}_{gg'} w^{(I)}_{ii'} w^{(G)}_{ag'} w^{(I)}_{bi'}
                    Y^{*}_{(t-1)g'i'}.                
    \end{equation*}
\fi
Although this covariance is always nonnegative, its magnitude depends on the susceptibility of the recipients, the overlap in their upstream sources, and the degree of transmission heterogeneity. 

Taken together, these derivations elucidate how added transmission heterogeneity yields a more flexible variance-covariance structure than the EE baseline without necessarily adjusting the process mean. 

%%%%%%%%%%%%%%%%%%%%%%%%%%%%%%%%%%%%%%%%%%%%%%%%%%%%%%%%%%%%%%%%%%%%%%%%%%%%%%%%

\subsection{Computation in the General Model}

\subsubsection{Generative Prior for Estimation of Age-Group Mixing Weights}
\label{ss:model-wstrata-prior}

An important estimand in doubly-stratified surveillance models is the between-age-groups mixing weights matrix $\bm{w}^{(I)}$. We estimate $\bm{w}^{(I)}$ indirectly by defining a generative model of a contact-rate matrix $\bm{C}$ for which $\bm{w}^{(I)}$ represents the mixing compositions (i.e. the column-wise normalization, $w^{(I)}_{ii'}∝C_{ii'}$). Because a contact-rate matrix is symmetric and elementwise-positive by definition, the crux of this method is selection of priors for the free (lower-triangular) elements of $\bm{C}$ that adhere to these constraints. Through a bespoke sampling scheme for this lower-triangle, we induce exchangeable gamma priors on the elements of $\bm{C}$ and thus exchangeable Dirichlet priors on the columns of $\bm{w}^{(I)}$. For the analyses presented among sections \Crefrange{s:simstudy}{s:inf-covid}, we use \mbox{$C_{ii'} \sim \mathrm{Gamma}(α_{ii'},\tfrac{1}{2I})$}. Hyperparameters $α_{ii} ≈ 4.3$ and $α_{ii'}≈1.3$ were chosen to loosely favor diagonal concentration, i.e. assortative mixing. \Cref{WA:clarify-wstrata-priors} details the sampling process and justifies these hyperparameter choices.

With no further constraints on the gamma hyperparameters $\{α_{ii'}\}$, the proposed method may flex to allow priors of varying informativeness. Though we regularize towards assortative mixing structures for our purposes, acceptable alternatives may include the use of a single global hyperparameter $α$ to favor homogeneous mixing, and use of unique $α_{ii'}$ for each $(i,i')$ to regularize towards a specific known contact matrix. In this way, our sampling scheme may incorporate user--directed shrinkage even with limited a priori knowledge of the mixing structure. 

%------------------------------------------------------------------------------%

\subsubsection{Lognormal Approximation to Gamma Offspring Distribution}

When fitting our proposed model in Stan, we sample $r_{tgi}$ from a non-centered lognormal approximation to the gamma offspring distribution seen in \Cref{eq:model-general-1}, specifically by matching the first two moments. While suboptimal, this is necessary for numerical stability. This is a limitation of our current software implementation rather than the model itself, but this is unlikely to change in the near future due to the difficulty of sampling from heterogeneous gamma mixtures. 

%%%%%%%%%%%%%%%%%%%%%%%%%%%%%%%%%%%%%%%%%%%%%%%%%%%%%%%%%%%%%%%%%%%%%%%%%%%%%%%%
%%%%%%%%%%%%%%%%%%%%%%%%%%%%%%%%%%%%%%%%%%%%%%%%%%%%%%%%%%%%%%%%%%%%%%%%%%%%%%%%

%%%%%%%%%%%%%%%%%%%%%%%%%%%%%%%%%%%%%%%%%%%%%%%%%%%%%%%%%%%%%%%%%%%%%%%%%%%%%%%%
%%%%%%%%%%%%%%%%%%%%%%%%%%%%%%%%%%%%%%%%%%%%%%%%%%%%%%%%%%%%%%%%%%%%%%%%%%%%%%%%

\section{Simulation Study}
\label{s:simstudy}

We conducted a simulation study to evaluate (i) recovery of key parameters and (ii) out-of-sample forecasting. The study is designed to isolate two modeling differences that motivate our proposed framework: explicit representation of transmission heterogeneity and estimation of between-ages mixing weights under a generative prior. 

%%%%%%%%%%%%%%%%%%%%%%%%%%%%%%%%%%%%%%%%%%%%%%%%%%%%%%%%%%%%%%%%%%%%%%%%%%%%%%%%
\subsection{Model Instance: Rare-Disease Specification} 
\label{ss:model-def-noro}

Here, we describe the specific instance of \Crefrange{eq:model-general-1}{eq:model-general-3} used in the simulation study. This specification is tailored to highly endemic, low-per-capita incidence observed on a reporting interval large relative to the infection period (e.g. weekly norovirus surveillance). In this regime, susceptible depletion is negligible over the analysis window, and the process can be modeled with an EE-type mean and an unbounded count likelihood.

With $E_{gi}$ denoting the size of subpopulation $(g,i)$, let $E_{∙∙}=\sum_{g,i}E_{gi}$ denote total population size. We begin by assuming $\hat{X}_{tgi}=E_{gi}$ and $\hat{Y}_{tgi}=Y^{*}_{tgi}$. We assume a log-additive decomposition of both the endemic term $δ_{tgi}$ and time-invariant susceptibility multiplier $ϕ_{gi}$, and we define offsets in terms of $E_{gi}$. Assuming observation time $t$ is given in weeks, time-varying effects of $δ_{gi}$ are limited to first-order harmonics with frequency $ω=\tfrac{2π}{52}$ and a short-term Christmas ``shock'' indicated by $x_{t}=\mathbbn{1}\{t∈\{1,52\}\}$. We write the decompositions as 
\ifbiomreferee
    % one-column / referee formatting
    \begin{equation}
    \setlength{\jot}{0mm}
    \begin{split}
        \log{δ_{tgi}}
            &= \log{\tfrac{E_{gi}}{E_{∙∙}}} + β_0 + β_{g}^{(G)} + β_{i}^{(I)} + β^{(S)}_{i}\sin(ωt) + β^{(C)}_{i}\cos(ωt) + β^{(\mathrm{xmas})}x_{t} \\
        \log{ϕ_{gi}}
            &= η^{(\mathrm{pop})}\log{\tfrac{E_{gi}}{E_{∙∙}}} + η_0 + η_{g}^{(G)} + η_{i}^{(I)}  
    \end{split}
    \label{eq:EE-delta-phi}
    \end{equation}
\else
    % two-column formatting
    \begin{equation}
    \setlength{\jot}{0mm}
    \begin{split}
        \log{δ_{tgi}}
            &= \log{\tfrac{E_{gi}}{E_{∙∙}}} + β_0 + β_{g}^{(G)} + β_{i}^{(I)} + β^{(S)}_{i}\sin(ωt)       \\ 
            &\hspace{24mm} + β^{(C)}_{i}\cos(ωt) + β^{(\mathrm{xmas})}x_{t} \\
        \log{ϕ_{gi}}
            &= η^{(\mathrm{pop})}\log{\tfrac{E_{gi}}{E_{∙∙}}} + η_0 + η_{g}^{(G)} + η_{i}^{(I)}  
    \end{split}
    \label{eq:EE-delta-phi}
    \end{equation}
\fi
where $β^{(G)}_{1}=β^{(I)}_{1}=0$ and $η^{(G)}_{1}=η^{(I)}_{1}=0$ are fixed for identifiability. Mixing weights $w^{(G)}_{gg'}w^{(I)}_{ii'}$ are treated as multiplicatively separable so that separate models may be considered for the spatial and between-age-groups mixing patterns. Mixing between areal units $g$ and $g'$ is given in terms of adjacency-order $o_{gg'}$ and unknown decay parameter $ρ$ as $w_{gg'}^{(\text{geog})} ∝ (1+o_{gg'})^{-ρ}$ before normalization. Mixing between age groups $i$ and $i'$ is given in terms of an unknown symmetric contact matrix $\bm{C}$ as $w_{ii'}^{(\text{strata})} ∝ C_{ii'}$ before normalization. With these pieces, we define realized infectiousness contributions $r_{tgi}$, infectious load $λ_{tgi}$, and conditional mean as in \Crefrange{eq:model-general-1}{eq:model-general-3}, with $R_{tgi}≡1$ fixed. We choose an identity-like link $g(\hat{X}_{tgi},λ_{tgi})=λ_{tgi}$. Using $\mathrm{NegBin}(μ,ψ)$ to denote to a negative binomial distribution with mean $μ$ and variance $μ(1+ψμ)$, our conditional likelihood is chosen to be
\begin{equation*}
    (Y^{*}_{tgi} \mid \bm{r}_{t}) \indepsim \mathrm{NegBin}\left(λ_{tgi}(\bm{r}_{t}), ψ \right).
\end{equation*}
The complete-data likelihood under these choices is 
\ifbiomreferee
    % one-column / referee formatting
    \begin{align*}
        \mathcal{L}(\bm{β}, \bm{η}, \bm{C}, ρ, ψ, θ) 
            =  \prod_{t} \prod_{g,i}
                    p_{\mathrm{NegBin}}(Y^{*}_{tgi} \mid \bm{r}_{t}, \bm{β}, \bm{η}, \bm{C}, ρ, ψ) ×
                    p_{\mathrm{Gamma}}(r_{tgi} \mid Y^{*}_{(t-1)gi}, θ). 
    \end{align*}
\else
    % two-column formatting
    \begin{align*}
        \mathcal{L}(\bm{β}, \bm{η}, \bm{C}, ρ, ψ, θ) 
            &= \prod_{t} \prod_{g,i}
                p_{\mathrm{NegBin}}(Y^{*}_{tgi} \mid \bm{r}_{t}, \bm{β}, \bm{η}, \bm{C}, ρ, ψ) \\
            &\qquad\qquad ×
                p_{\mathrm{Gamma}}(r_{tgi} \mid Y^{*}_{(t-1)gi}, θ).
    \end{align*}
\fi

%%%%%%%%%%%%%%%%%%%%%%%%%%%%%%%%%%%%%%%%%%%%%%%%%%%%%%%%%%%%%%%%%%%%%%%%%%%%%%%%
\subsection{Models Under Investigation}

Our simulation considers three models. The ``reduced'' model, introduced in \citet[eq.~4.1]{2017-Meyer-Held-Social-Contact}, serves as our primary point of comparison. This is a doubly-stratified EE model which treats between-ages mixing weights as a low-dimensional deformation of an external contact matrix. Specifically, Meyer and Held assume $w_{ii'}^{(I)} ∝ [\bm{Ω}\bm{Λ}^{κ}\bm{Ω}^{-1}]_{ii'}$, where $\bm{Ω}$ and $\bm{Λ}$ denote the matrices of eigenvectors and eigenvalues obtained from a known (and pre-normalized) contact matrix. Here, $κ$ is an unknown parameter that interpolates between assortative mixing ($κ\downarrow 0$) and origin-invariant mixing ($κ\uparrow ∞$). 

The ``full'' model is described in the previous subsection (\Cref{ss:model-def-noro}). We've intentionally defined the peripheral components of this model (such as the definitions in \cref{eq:EE-delta-phi}) to match that of the Meyer-Held model. The terms ``full'' and ``reduced'' are used not to diminish the value of the Meyer-Held comparator, but to emphasize that our proposed model subsumes it as a special case. Our model instance differs from the Meyer-Held model \textit{only} in (i) the assumed structure used to estimate age-group mixing weights and (ii) the representation of transmission heterogeneity. This way, the full model may be recognized as the application of our methodology to the Meyer-Held model (up to the simplification $ψ_{i}≡ψ$ made for parsimony of our simulation study).   

The third, ``naive'', model serves as a baseline for forecasting performance in settings with strong endemic seasonality but non-negligible epidemic feedback. This naive model bears the same endemic structure $δ_{tgi}$ as the full and reduced models but drops the autoregressive (``epidemic'') component entirely (fixing $ϕ_{tgi}=0$), making it ``naive'' to the data generating process' natural self-excitation. 

%%%%%%%%%%%%%%%%%%%%%%%%%%%%%%%%%%%%%%%%%%%%%%%%%%%%%%%%%%%%%%%%%%%%%%%%%%%%%%%%
\subsection{Data-Generating Process}
\label{ss:sim-dgp}

Data were generated according to the full model under conditions mimicking the intended application regime of the Meyer-Held model: incident counts exhibiting high endemicity, strong seasonal dynamics, and low incidence per capita. We set the panel dimensions equal to those of the Berlin norovirus data analyzed by Meyer and Held. This includes $G=12$ areal units, $I=6$ age groups, and $T=208$ weekly observations for model fitting, followed by an additional contiguous testing period of $H=52$ weeks for out-of-sample forecasting evaluation. With the exception of dispersion parameters $θ$ and $ψ$ and age-group mixing weights $\bm{w}^{(I)}$, all parameters were fixed a priori to values chosen to imitate the aforementioned Berlin norovirus dataset. In particular, we generated data using weights $w^{(I)}_{ii'} ∝ C_{ii'}$ for a \textit{known} contact matrix $\bm{C}$. This is consistent with the eigen-deformation assumption of the Meyer-Held model (where $κ=1$ is the truth), meaning that both the full and reduced models were fit without misspecification of the mixing structure. 

Across datasets, we varied the parameters $θ$ (latent-infectiousness-level dispersion) and $ψ$ (observation-level dispersion) responsible for driving tail behavior and predictive uncertainty. Specifically, we defined 16 scenarios for model comparison by the $4×4$ grid product $(θ,ψ)∈\{0.05,5,15,40\}×\{0.05,0.5,1,3\}$. Though data were generated under the full model, scenarios featuring $θ$ close to the boundary of the parameter space were included to assess our proposed model's performance when $θ$ is practically non-identifiable. For each scenario, we generated 200 independent datasets to which all three models were fit. 

%%%%%%%%%%%%%%%%%%%%%%%%%%%%%%%%%%%%%%%%%%%%%%%%%%%%%%%%%%%%%%%%%%%%%%%%%%%%%%%%
\subsection{Model Fitting Procedure}
\label{ss:sim-model-fit-proc}

Models were fit in Stan using no U-turn sampling (NUTS), each fit consisting of 4 chains of 1500 warmup iterations and 1500 retained draws, thus $6000$ posterior draws per model fit. Fitting the reduced model in Stan rather than utilizing the \texttt{hhh4contacts} R package associated with \citet{2017-Meyer-Held-Social-Contact} is necessary for a coherent model comparison (the \texttt{hhh4contacts} implementation uses a frequentist profiled-likelihood approach). Given the computational expense of fitting the full and reduced models, all fits were conducted on the Oregon State University College of Engineering's (OSU COE) high-performance computing cluster. 

Across all three models, parameters shared in common (e.g., endemic parameters $\bm\beta$) were assigned the same priors, as listed in \Cref{WA:clarify-other-priors}. Priors for between-ages mixing parameters $\bm{w}^{(I)}$ are described in \Cref{ss:model-wstrata-prior} for the full model. For the reduced model, we assume $\mathrm{log}(κ) \sim \Normal{0,\tfrac{3}{4}}$ with the rationale that it (i) is approximately centered at $κ=1$, the natural null hypothesis of no eigen-power adjustment; and, (ii) places minimal mass on extreme values (say, $κ<0.1$ or $κ>5$) for which the resulting $\bm{w}^{(I)}$ matrix would be practically non-identifiable.

To preserve a single workflow across all 16 simulation scenarios, we used common weakly-informative priors for $θ$ and $ψ$ rather than scenario-specific priors. These priors were chosen to concentrate on the regimes in which $θ$ and $ψ$ are practically identifiable, regularizing away from the boundary of the parameter space, where latent- and observation-level dispersion become difficult to distinguish and posterior geometry is most challenging for HMC. Accordingly, scenarios with $θ=0.05$ and/or $ψ=0.05$ should be interpreted as deliberate stress tests of boundary robustness under moderate prior misspecification. For the data analyses performed in \Cref{s:inf-noro,s:inf-covid}, we instead use priors amenable to near-boundary estimation of $θ$ and $ψ$ and account for the associated computational difficulties by increasing the number of HMC chains and iterations. 

%%%%%%%%%%%%%%%%%%%%%%%%%%%%%%%%%%%%%%%%%%%%%%%%%%%%%%%%%%%%%%%%%%%%%%%%%%%%%%%%
\subsection{Evaluation Criteria \& Simulation Results}

We first evaluate the full and reduced models on the basis of parameter recovery. Specifically, we report these models' empirical coverage of 90\% credible intervals across replicate datasets. Given the high-dimensionality of the parameter space, and given the lack of commonality of select parameters among the models (e.g. eigen-power $κ$), we limit our scope to the key dispersion parameters $θ$ and $ψ$ and select between-ages mixing parameters $w^{(I)}_{ii'}$.

\Cref{tbl:coverage-psi} details empirical coverage of 90\% credible intervals for observation-level dispersion $ψ$ in both models. Though coverage is similar when transmission heterogeneity is negligible ($θ=0.05$), we find that coverage for the reduced model decreases in $θ$ and increases in $ψ$. Coverage for the full model is stable at approximately the desired 90\% level, except for the $ψ=0.05$ scenarios. \Cref{tbl:coverage-theta} details empirical coverage for latent dispersion $θ$, present only in the full model. Coverage was expectedly poor for a true $θ=0.05$, stable at the desired 90\% level for moderate $θ$, and diminished as $θ$ increased. Undercoverage for the $θ=0.05$ and $ψ=0.05$ scenarios reflects tension between the near-boundary truth and the (intentionally) misspecified prior rather than a general failure of the proposed model in true equidispersion regimes. Undercoverage for the $θ=40$ scenarios is likely explained by the moderate negative correlation observed between $θ$ and $ψ$ in the posterior.

\begin{table}[b!]
    \addtolength{\tabcolsep}{-0.1em}
    \linespread{1.3}\selectfont
    \centering
    \begin{tabular}{ccccccccc} \hline
        \multicolumn{1}{c}{ } & 
            \multicolumn{2}{c}{\ensuremath{\theta=0.05}} & 
            \multicolumn{2}{c}{\ensuremath{\theta=5}} & 
            \multicolumn{2}{c}{\ensuremath{\theta=15}} & 
            \multicolumn{2}{c}{\ensuremath{\theta=40}} \\
        \cline{2-3} \cline{4-5} \cline{6-7} \cline{8-9}
        \ensuremath{\psi} & MH & F & MH & F & MH & F & MH & F\\ \hline
        0.05 & 0.81 & 0.92 & 0.00 & 0.72 & 0.00 & 0.53 & 0.00 & 0.66\\ \hline
        0.50 & 0.90 & 0.84 & 0.09 & 0.88 & 0.00 & 0.85 & 0.00 & 0.86\\ \hline
        1.00 & 0.90 & 0.86 & 0.27 & 0.88 & 0.00 & 0.86 & 0.00 & 0.85\\ \hline
        3.00 & 0.87 & 0.85 & 0.65 & 0.89 & 0.14 & 0.87 & 0.00 & 0.84\\ \hline
    \end{tabular}
    % \vspace{1mm}
    \caption{Empirical coverage of observation-level dispersion $ψ$ for full (F) and reduced (R) models. Each observed coverage proportion is an average over 200 synthetic datasets, rounded to 2 digits}
    \label{tbl:coverage-psi}
\end{table}

\begin{table}[b!]
    \addtolength{\tabcolsep}{-0.1em}
    \linespread{1.3}\selectfont
    \centering
    \begin{tabular}{ccccc} \hline
        \ensuremath{\psi} & \ensuremath{\theta=0.05} & \ensuremath{\theta=5} & \ensuremath{\theta=15} & \ensuremath{\theta=40} \\ \hline
        0.05 & 0.02 & 0.93 & 0.50 & 0.02 \\ \hline
        0.50 & 0.01 & 0.94 & 0.61 & 0.09 \\ \hline
        1.00 & 0.00 & 0.91 & 0.72 & 0.28 \\ \hline
        3.00 & 0.00 & 0.94 & 0.88 & 0.74 \\
    \hline
    \end{tabular}
    % \vspace{1mm}
    \caption{Empirical coverage of latent-level dispersion $θ$ for the full model. Each observed coverage proportion is an average over 200 synthetic datasets, rounded to 2 digits.}
    \label{tbl:coverage-theta}
\end{table}

\begin{figure}[!ht]
    % \hspace{-7mm}
    % \centering
    \linespread{1.3}\selectfont    
    \includegraphics[width=\linewidth]{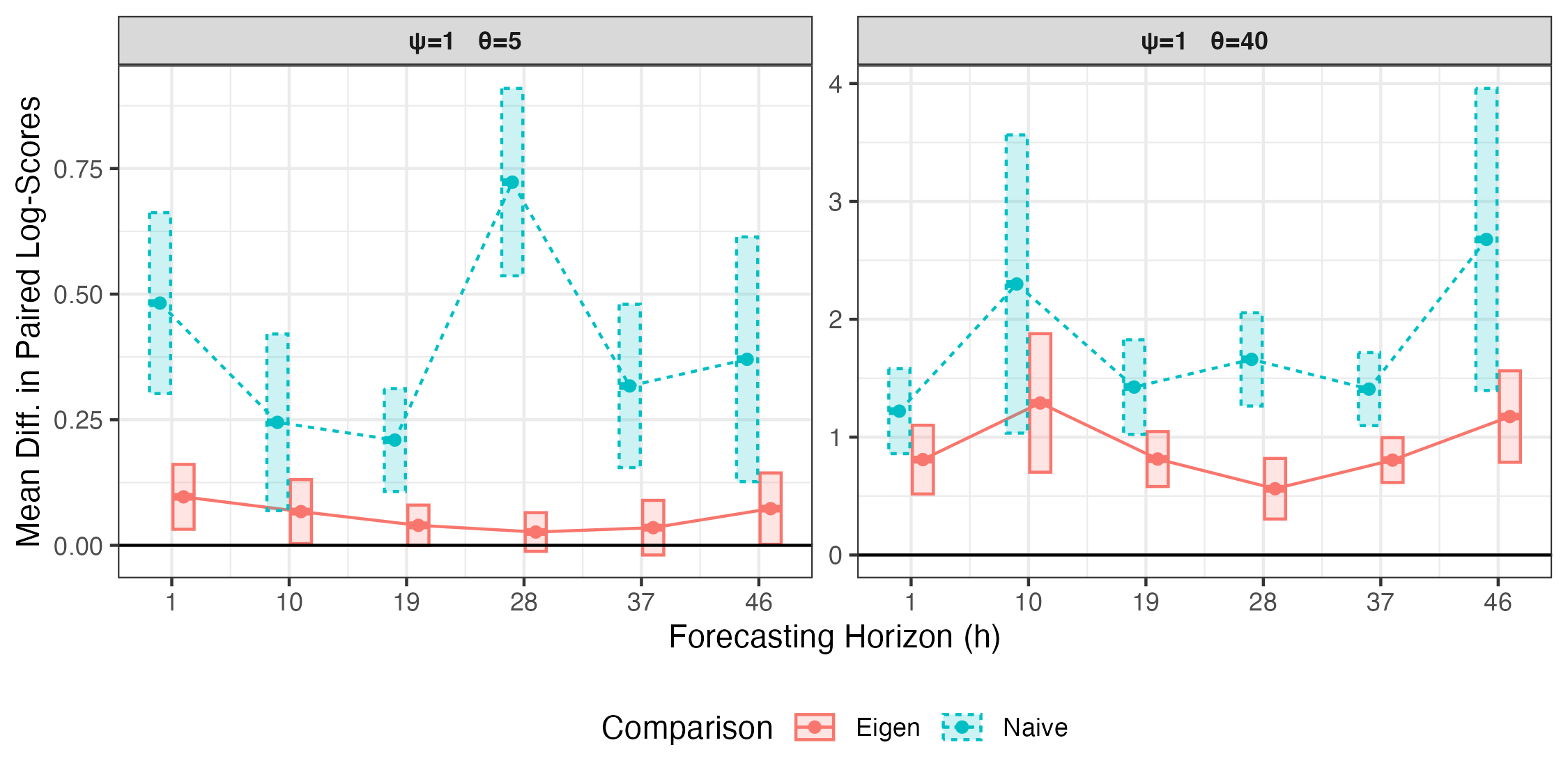}
    \caption{Mean difference in paired log-scores. Intervals for the full-vs-naive comparison are given in blue (dashed lines), and intervals for the full-vs-reduced comparison are given in red (solid lines). Computation of these intervals is described in \Cref{WA:clarify-log-score}. Intervals are offset (``jittered'') along the $x$-axis for visual separability.}
    \label{plot:simstudy-logscores-small}
    \vspace{5mm}
\end{figure}

For each model, we also evaluate accuracy and precision of out-of-sample forecasts in tandem using the logarithmic score \citep{2007-Gneiting-Raftery-Scoring-Rules}. Specifically, we score the $h$-step-ahead posterior predictive distributions obtained from each model against the true value $\bm{Y}^{*(d)}_{T+h}$ generated alongside each dataset $d$. This forecasting target is multivariate (a $G×I$ matrix), so we obtain an estimate of a single log-score ${\mathrm{LS}}^{(d)}_{m,h}$ for each $(d,h)$ for each model $m$. We then aggregate over models as paired-differences and report averages of these paired-differences over datasets. A detailed description of our evaluation procedure is given in \Cref{WA:clarify-log-score}.

For two chosen scenarios, \Cref{plot:simstudy-logscores-small} depicts these average paired-differences alongside 95\% intervals estimated as $\pm 1.96\cdot\mathrm{SE}$. Corresponding plots for the remaining scenarios are included in the Web Appendix as \Cref{plot:simstudy-logscores-full}. For the scenarios in which transmission heterogeneity is negligible ($θ=0.05$), we see near-identical predictive performance of the full and reduced model. This holds even for long prediction horizons, suggesting relative unimportance of the mixing structure in population-level predictive performance. Unsurprisingly, the predictive advantage of the full model over the reduced and naive models increases with $θ$. The worst-case scenario for a classical EE model is presented in the top-right plot ($ψ=0.05$, $θ=40$), in which the full model places around $3.6-13.2×$ more mass on the truth than does the reduced model. The predictive superiority of the full model seems to diminish as observation-level dispersion $ψ$ increases. Though this suggests competence of the EE model in the presence of transmission heterogeneity (so long as $ψ$ is not too small), this could also have to do with the relative importance of the mean structure (which these models have in common) over the variance-covariance structure when using log-score to assess population-level predictions.  

Standard Bayesian diagnostic summaries were computed to assess the validity of these results. For each dataset, we obtained the maximum $\hat{R}$ statistic among all parameters. While the reduced and naive model fits saw $\mathrm{max}(\hat{R})<1.02$ universally, around $4\%$ of full model fits saw \mbox{$\mathrm{max}(\hat{R})>1.02$}. We believe these failures were the result of single chains over-adapting to funnel geometry near the boundary of the parameter space during warmup, and could be addressed by tuning certain priors or sampling hyperparameters. 

%%%%%%%%%%%%%%%%%%%%%%%%%%%%%%%%%%%%%%%%%%%%%%%%%%%%%%%%%%%%%%%%%%%%%%%%%%%%%%%%
%%%%%%%%%%%%%%%%%%%%%%%%%%%%%%%%%%%%%%%%%%%%%%%%%%%%%%%%%%%%%%%%%%%%%%%%%%%%%%%%

%%%%%%%%%%%%%%%%%%%%%%%%%%%%%%%%%%%%%%%%%%%%%%%%%%%%%%%%%%%%%%%%%%%%%%%%%%%%%%%%
%%%%%%%%%%%%%%%%%%%%%%%%%%%%%%%%%%%%%%%%%%%%%%%%%%%%%%%%%%%%%%%%%%%%%%%%%%%%%%%%

\section{Inference: Norovirus Incidence in Berlin}
\label{s:inf-noro}

Here, we detail the results of applying the rare-disease model instance from \Cref{ss:model-def-noro} to weekly norovirus gastroenteritis counts in Berlin, from 2011 to 2015. This dataset was previously analyzed by \citet{2017-Meyer-Held-Social-Contact}. We demonstrate the benefits of our model over existing frameworks, and we find evidence of substantial transmission heterogeneity. 

%%%%%%%%%%%%%%%%%%%%%%%%%%%%%%%%%%%%%%%%%%%%%%%%%%%%%%%%%%%%%%%%%%%%%%%%%%%%%%%%
\subsection{Data Description \& Model Fitting Procedure}
\label{ss:noro-data-description}

We used the dataset as aggregated by \citeauthor{2017-Meyer-Held-Social-Contact}, comprising $T=208$ weekly observations across $G=12$ administrative districts and $I=6$ age-groups. These data were accessed through the accompanying \texttt{hhh4contacts} R package authored by Meyer and Held.

We fit two Stan models to this dataset: (i) the ``full'' model described in \Cref{ss:model-def-noro}, corresponding to our proposed latent-infectiousness formulation with generative prior estimation of $\bm{w}^{(I)}$; and (ii) the ``reduced'' model of \Cref{ss:model-def-noro}, corresponding to a Bayesian implementation of the Meyer-Held eigen-deformation model. Each Stan fit used 16 chains with $2{,}000$ warmup iterations and $2{,}000$ retained draws. For comparison, we also re-fit the Meyer-Held model using \texttt{hhh4contacts}.

Importantly, the priors on $θ$ and $ψ$ deviate from those used in the simulation study. We used priors $θ,ψ\sim\mathrm{HalfCauchy}(\mathrm{scale}=1)$, which yield reliable estimation for true $θ≈0$ and/or $ψ≈0$. Sampler diagnostics passed routine inspection, including $\mathrm{max}(\hat{R})<1.01$ for both models. 

%%%%%%%%%%%%%%%%%%%%%%%%%%%%%%%%%%%%%%%%%%%%%%%%%%%%%%%%%%%%%%%%%%%%%%%%%%%%%%%%
\subsection{Berlin Norovirus Results}

\Cref{tbl:inf-noro-results} summarizes posterior inference for select mixing and dispersion parameters. Two observations are immediate. First, our model reproduces the broad conclusions about age-mixing made in \citet{2017-Meyer-Held-Social-Contact} despite using substantially less information. Estimates for diagonal entries $w^{(I)}_{ii}$ are large, consistent with their finding of strong assortative mixing. At the same time, the full model assigns appreciably-wider uncertainty to $\bm{w}^{(I)}$ and is not restricted to a one-dimensional manifold, as is the case with the Meyer-Held model. This demonstrates the viability of inferring age mixing structure from surveillance data alone.

Second, the data support non-negligible transmission heterogeneity. The posterior for $θ$ concentrates away from the EE limit $θ=0$, with a 90\% credible interval $(5.14, 9.27)$. Correspondingly, the full model estimates a smaller observation-level dispersion $ψ$ than the reduced model. This highlights another advantage of our model: rather than attributing all extra-Poisson variation to observational noise, overdispersion is partitioned into both observational noise and variation in latent individual-level infectiousness.  

Our second finding is qualitatively consistent with \citet{2020-Zelner-Transmission-Heterogeneity}, who concluded that transmission heterogeneity underpinned the ``source-and-sink'' dynamics observed in their own norovirus outbreak dataset. Though our result lacks some of the mechanistic detail (e.g. an explicit incubation period) present in the \citeauthor{2020-Zelner-Transmission-Heterogeneity} analysis, it shows that evidence for the same general phenomenon remains visible in broader surveillance data settings.

Taken together, the norovirus application suggests that the main conclusions of the Meyer-Held analysis are recoverable under looser modeling and data source constraints, while still revealing evidence of transmission heterogeneity absent from the classical EE framework.

\begin{table}[b!]
\addtolength{\tabcolsep}{-0.22em}
\def\arraystretch{1.3}
\linespread{1.3}\selectfont
\centering

\begin{tabular}{crrr} \hline
\multicolumn{1}{c}{Var} & \multicolumn{1}{c}{Full} & \multicolumn{1}{c}{MH} & \multicolumn{1}{c}{\texttt{hhh4contacts}} \\ \hline
$w^{(I)}_{11}$  &0.60 (0.44--0.74)&0.50 (0.41--0.60)&0.49 (0.40--0.60)\\ \hline
$w^{(I)}_{22}$  &0.40 (0.18--0.65)&0.64 (0.57--0.72)&0.64 (0.56--0.73)\\ \hline
$w^{(I)}_{12}$  &0.07 (0.01--0.19)&0.04 (0.03--0.04)&0.04 (0.03--0.04)\\ \hline
$\psi$          &0.19 (0.16--0.22)&0.31 (0.29--0.34)&0.31 (0.29--0.32)\\ \hline
$\theta$        &6.89 (5.14--9.27)&--&--\\ \hline
\end{tabular}
\caption{Marginal point estimates and intervals for key model parameters obtained from Norovirus dataset. Full and MH report 90\% equal-tailed credible intervals. Full and MH point-estimates are medians for $ψ$ and $θ$ and posterior means for $w^{(I)}_{ii'}$. \texttt{hhh4contacts} reports MLE and 90\% Wald confidence intervals. %Note that joint credible sets for $w^{(I)}_{ii'}$ actually live on a 1D subspace for the MH and \texttt{hhh4contacts} results.}
}
\label{tbl:inf-noro-results}
\end{table}

%%%%%%%%%%%%%%%%%%%%%%%%%%%%%%%%%%%%%%%%%%%%%%%%%%%%%%%%%%%%%%%%%%%%%%%%%%%%%%%%
%%%%%%%%%%%%%%%%%%%%%%%%%%%%%%%%%%%%%%%%%%%%%%%%%%%%%%%%%%%%%%%%%%%%%%%%%%%%%%%%

 %%%%%%%%%%%%%%%%%%%%%%%%%%%%%%%%%%%%%%%%%%%%%%%%%%%%%%%%%%%%%%%%%%%%%%%%%%%%%%%%
%%%%%%%%%%%%%%%%%%%%%%%%%%%%%%%%%%%%%%%%%%%%%%%%%%%%%%%%%%%%%%%%%%%%%%%%%%%%%%%%

\section{Inference: COVID-19 Incidence in Michigan}
\label{s:inf-covid}

We now present the results of an application of our model to COVID-19 cases in Michigan during the first year of the pandemic. These data are characterized by substantial non-seasonal temporal variation; non-negligible susceptible depletion for some age-by-geography cells; and nonexistent a priori knowledge of age-group mixing patterns. Our aim in this section is twofold: (i) to introduce a model instance adapted to outbreak-driven incidence for which the typical rare disease assumptions do not hold, and (ii) to use this adapted model for inference on a previously unanalyzed doubly-stratified surveillance dataset.

%%%%%%%%%%%%%%%%%%%%%%%%%%%%%%%%%%%%%%%%%%%%%%%%%%%%%%%%%%%%%%%%%%%%%%%%%%%%%%%%
\subsection{Data Description}

The dataset consists of COVID-19 case counts in Michigan aggregated by Public Use Microdata Area (PUMA) and age group over a contiguous period, March 2020 to January 2021. The analysis panel contains $T=45$ weekly observation times, $G=66$ PUMAs, and $I=6$ age groups binned as 0-9, 10-19, 20-29, 30-39, 40-59, and 60+. We exclude the MI upper peninsula, as its connectivity structure is qualitatively different from the contiguous lower-peninsula PUMAs. Incidence data were provided by the Michigan Department of Health and Human Services (MDHHS), and population counts $E_{gi}$ were obtained from the 2020 census through the \texttt{tidycensus} R package \citep{2025-tidycensus}.  

This application differs from the Berlin norovirus setting in three respects. First, incidence is concentrated in waves coinciding with the introduction and relaxation of stay-at-home orders, rather than fluctuating around a persistent endemic baseline. Second, incidence is sufficiently high that the upper bound \mbox{$Y^{*}_{tgi}≤X_{tgi}$} cannot be ignored, particularly for smaller subpopulations. Third, incidence is not readily augmented with external between-ages contact data, as no such data exist for this population. These features motivate an ``outbreak-driven'' specification featuring explicit components for transmission-heterogeneity and susceptible depletion, as well as direct estimation of age-group mixing weights.

%%%%%%%%%%%%%%%%%%%%%%%%%%%%%%%%%%%%%%%%%%%%%%%%%%%%%%%%%%%%%%%%%%%%%%%%%%%%%%%%
\subsection{Model Instance: Outbreak-Driven Specification}
\label{ss:model-def-covid}

Relative to the rare-disease specification of \Cref{ss:model-def-noro}, the present model modifies both the conditional likelihood and the approximations of susceptible population size and infectious prevalence. We retain the general latent-infectiousness framework of \Crefrange{eq:model-general-1}{eq:model-general-3}, but replace the assumptions of negligible depletion and one-period infectiousness with estimators tailored to an outbreak setting.

We model depletion of the susceptible pool as permanent over the analysis window. To allow infectiousness to persist beyond a single observation interval, we estimate infectious prevalence through a distributed lag model parameterized by decay rate $γ>0$. Formally, 
\ifbiomreferee
    \begin{align*}
        \hat{X}_{tgi} &= \left[ E_{gi} - \sum_{d=1}^{t-1} Y^{*}_{(t-d)gi} \right] &
        \! \hat{Y}_{tgi} &= \left[ \sum_{d=1}^{t-1} e^{-γ(d-1)} Y^{*}_{(t-d)gi} \right].
    \end{align*}
\else
    \begin{align*}
        \hat{X}_{tgi} &= \left[ E_{gi} - \sum_{d=1}^{t-1} Y^{*}_{(t-d)gi} \right] &
        \! \hat{Y}_{tgi} &= \left[ \sum_{d=1}^{t-1} e^{-γ(d-1)} Y^{*}_{(t-d)gi} \right].
    \end{align*}
\fi 

Conditional on $\hat{Y}_{tgi}$ above, realized infectiousness is described as in \Cref{eq:model-general-1}, with mean $R_{t-1}α_{i}\hat{Y}_{tgi}$ and dispersion parameter $θ$. Here, $R_{t-1}$ is a finite-sample analog of the time-varying reproduction number, and $θ$ governs transmission heterogeneity as in the general model. Unique to this model, $α_{i}$ functions similar to the subpopulation-specific susceptibility parameter $ϕ_{tgi}$ seen in \Cref{eq:model-general-2}: it describes age-group-specific ``activity,'' i.e. contact volume. By assuming that $α_{i}$ is derived from the same contact matrix $\bm{C}$ as our age-group mixing weights, we aim to extract additional signal for $\bm{C}$ itself.

Mixing is again assumed separable across geography and age group, written $w^{(G)}_{gg'} w^{(I)}_{ii'}$. For between-age-group mixing, we use the same generative model construction described in \Cref{ss:model-wstrata-prior} to estimate $\bm{w}^{(I)}$ as a normalized contact matrix, but with the additional assumption that the $α_{i}$ now describe the column-totals of $\bm{C}$.  

Geographic mixing weights are modeled \mbox{$w^{(G)}_{gg'} ∝ τ_{g} D^{-ρ_{g'}}_{gg'}$}, where $τ_{g}>0$ represents destination-specific attractiveness and $ρ_{g'}>0$ represents source-specific distance-decay rate. The $ρ_{g'}$ are estimated as random effects. The distance matrix $\bm{D}$ is fixed a priori and described in \Cref{WA:puma-distmat}. This parameterization allows both the propensity to export infections and the spatial distribution of recipients to vary across areal units.

Given infectiousness contributions $\bm{r}_t$, we define the infection probability in cell $(g,i)$ by
\begin{equation}
    p_{tgi}
        = 1 - \mathrm{exp}\left\{ -\left( δ_{gi} + \frac{1}{E_{gi}} 
                \sum_{g',i'} w^{(G)}_{gg'} w^{(I)}_{ii'} r_{tg'i'} 
            \right) \right\},
\end{equation}
where $δ_{gi}>0$ is a cell-specific endemic hazard estimated outright. Incidence is then modeled with a beta-binomial conditional likelihood, $(Y^*_{tgi}\mid \hat{X}_{tgi}, \bm{r}_t) \indepsim \mathrm{BetaBinom}$ with size $\hat{X}_{tgi}$, average probability $p_{tgi}$, and precision parameter $k>0$. Note that, with variance $Np(1-p)\left(\frac{k+N}{k+1}\right)$, the limit $k\uparrow ∞$ corresponds to a classic equidispersed binomial distribution (see \Cref{WA:beta-binom}). 

This choice preserves the natural upper bound $Y^{*}_{tgi} ≤ \hat{X}_{tgi}$ while still allowing extra-binomial variation. In contrast to the negative-binomial likelihood used in \Cref{ss:model-def-noro}, the present specification is therefore better aligned with epidemic periods in which incidence may constitute a non-trivial fraction of the susceptible population.

Combined, these changes produce a model instance whose mean structure is still interpretable through ``endemic'' and ``epidemic'' contributions, but whose variance structure and temporal mechanics are more suitable for outbreak-driven incidence. The rare-disease and outbreak-driven specifications should thus be viewed as two disease-specific instantiations of the same general latent-infectiousness framework, differing primarily in their consideration of dispersion, infection duration, and re-infection mechanics.

%%%%%%%%%%%%%%%%%%%%%%%%%%%%%%%%%%%%%%%%%%%%%%%%%%%%%%%%%%%%%%%%%%%%%%%%%%%%%%%%
\subsection{COVID-19 Model Fitting Procedure}

We fit this model in Stan using 16 chains with $2{,}000$ warmup iterations and $2{,}000$ retained draws. Priors for $\bm{w}^{(I)}$ are described in \Cref{ss:model-wstrata-prior} and \Cref{WA:clarify-wstrata-priors}, and priors for nuisance parameters are described in \Cref{WA:clarify-covid-priors}. For our dispersion parameters, we use weakly-informative priors \mbox{$θ\sim\mathrm{HalfCauchy}(\mathrm{scale}=5)$} and \mbox{$k\sim\mathrm{HalfCauchy}(\mathrm{scale}=10^4)$}.

%%%%%%%%%%%%%%%%%%%%%%%%%%%%%%%%%%%%%%%%%%%%%%%%%%%%%%%%%%%%%%%%%%%%%%%%%%%%%%%%
\subsection{COVID-19 Model Results}

\begin{table}[b!]
\addtolength{\tabcolsep}{-0.22em}
\def\arraystretch{1.3}
\linespread{1.3}\selectfont
\centering

\begin{tabular}{crr} \hline
\multicolumn{1}{c}{Param} & \multicolumn{1}{c}{Prior} & \multicolumn{1}{c}{Posterior} \\ \hline
$w^{(I)}_{11}$ &0.40 (0.18--0.64)&0.92 (0.90--0.94)\\ \hline
$w^{(I)}_{22}$ &0.40 (0.18--0.65)&0.82 (0.80--0.83)\\ \hline
$w^{(I)}_{12}$ &0.12 (0.01--0.31)&0.03 (0.02--0.05)\\ \hline
$\gamma$       &0.99 (0.08--13.07)&1.64 (1.49--1.81)\\ \hline
$\theta$       &5.07 (0.40--62.38)&15.67 (14.54--16.89)\\ \hline
$k/10^{4}$     &1.00 (0.08--12.60)&7.02 (6.24--7.98)\\ \hline
\end{tabular}

% \vspace{1mm}
\caption{Marginal point estimates and 90\% equal-tailed intervals for key model parameters obtained from COVID-19 dataset. Posterior mean are given for $w^{(I)}_{ii'}$ and posterior medians for $γ$, $θ$, and $k/10^{4}$.}
\label{tbl:inf-covid-results}
\end{table}
\begin{figure}[!ht]
    \hspace{-9mm}
    \linespread{1.3}\selectfont    
    \includegraphics[width=1.1\linewidth]{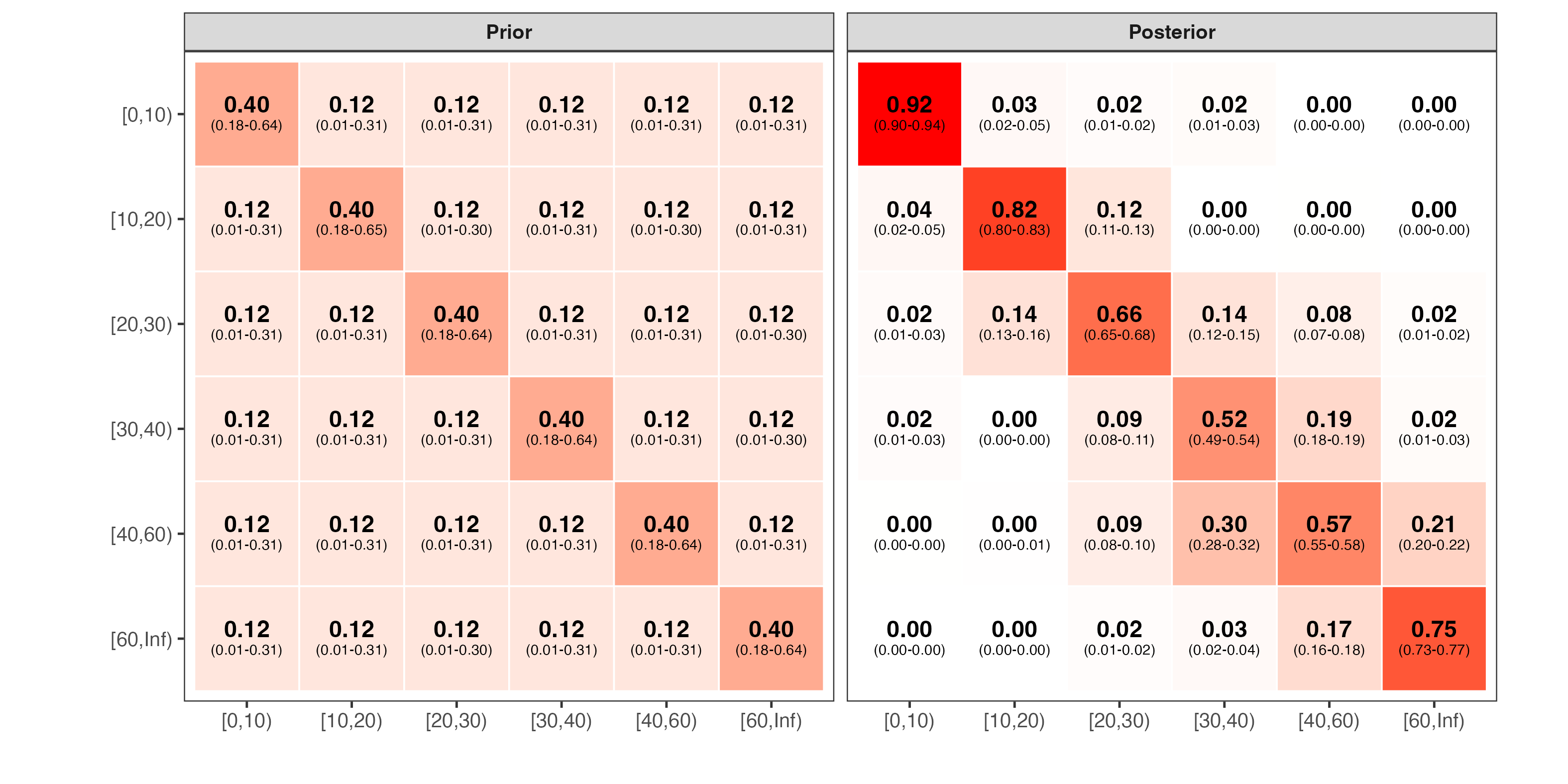}
    \caption{Prior and posterior means (and 90\% equal-tailed intervals) for elements of $\bm{w}^{(I)}$ from the COVID-19 model fit. Origin age-groups are depicted on the x-axis and recipients on the y-axis. Both matrices are column-stochastic.}
    \label{plot:inf-covid-wstrata-estm}
    \vspace{5mm}
\end{figure}

\Cref{tbl:inf-covid-results} summarizes posterior inference for select age-mixing and dispersion parameters, and \Cref{plot:inf-covid-wstrata-estm} summarizes the prior and posterior distributions for $\bm{w}^{(I)}$. Three findings are most notable. First, the posterior estimates concentrate on a regime in which the latent-level dispersion parameter $θ$ and the observation-level dispersion parameter $k$ contribute unique information about observed extra-binomial dispersion. With a posterior 90\% credible interval $(14.54,16.89)$ for $θ$, we have strong evidence for the presence of transmission heterogeneity. For the beta-binomial precision parameter $k$, the 90\% credible interval $(6.0\cdot 10^4, 7.6\cdot 10^4)$ leads to a factor of 1.14--1.76$×$ greater conditional variance compared to an equidispersed binomial. Low posterior correlation $\Cor{θ,k}≈0.01$ suggests that the two dispersion mechanisms are not mutually-redundant; so, the epidemic component appears to require substantial variability in latent infectious contribution, even after accounting for observation-level overdispersion, susceptible depletion, and mixing structure. 

Second, the data provide strong evidence of assortative age-group mixing. Though the prior on $\bm{w}^{(I)}$ regularized towards mild assortativity, the posterior distribution showed strong signal favoring \textit{extreme} within-groups-dominated transmission dynamics. This is reflected in \Cref{tbl:inf-covid-results} by large posterior means on diagonal entries, such as $\hat{w}^{(I)}_{11}≈0.91$. As each column of $\bm{w}^{(I)}$ is a composition, the off-diagonal entries show posterior updating towards 0. 

Third, the posterior for the prevalence-decay rate $γ$ indicates that infectious contribution is concentrated in the most recent lags. With posterior 90\% credible interval $(1.49,1.81)$, estimated prevalence is clearly dominated by  lag-1 and lag-2 incidence. This behavior is consistent with a short-memory epidemic process at the weekly resolution of the data.

Overall, these results suggest that the Michigan COVID-19 data contain substantial information about age-assortative mixing and transmission heterogeneity. The fitted model attributes the observed epidemic dynamics primarily to structured epidemic feedback and latent variability in realized infectiousness, rather than to large observation-level noise.

%%%%%%%%%%%%%%%%%%%%%%%%%%%%%%%%%%%%%%%%%%%%%%%%%%%%%%%%%%%%%%%%%%%%%%%%%%%%%%%%
%%%%%%%%%%%%%%%%%%%%%%%%%%%%%%%%%%%%%%%%%%%%%%%%%%%%%%%%%%%%%%%%%%%%%%%%%%%%%%%%
%%%%%%%%%%%%%%%%%%%%%%%%%%%%%%%%%%%%%%%%%%%%%%%%%%%%%%%%%%%%%%%%%%%%%%%%%%%%%%%%
%%%%%%%%%%%%%%%%%%%%%%%%%%%%%%%%%%%%%%%%%%%%%%%%%%%%%%%%%%%%%%%%%%%%%%%%%%%%%%%%

\section{Discussion} 
\label{s:discuss}

We introduced a latent-infectiousness generalization of the endemic-epidemic framework for doubly-stratified surveillance data. Our proposal preserves the familiar decomposition of a linear predictor into endemic and epidemic components, but advances the existing EE framework in three substantive ways. First, it permits a broader range of variance-covariance structures by way of a ``transmission heterogeneity'' latent variable. Second, it permits estimation of strata (e.g. age-group) mixing weights under looser structural assumptions than the eigen-deformation approach of \citet{2017-Meyer-Held-Social-Contact}. Third, it permits explicit representation of susceptible pool size and disease prevalence, extending its use cases to non-rare-disease contexts. The analyses considered here demonstrate the practical value of these extensions. 
The simulation study revealed improved predictive performance when transmission heterogeneity is present; the norovirus application recovered the main age-mixing conclusions of Meyer and Held under our more flexible inferential scheme; and, the COVID-19 application reaffirms the identifiability of the new model components in a relevant outbreak-driven application context.

While much of our model's flexibility stems from the chosen Bayesian workflow, the principal limitation of our methodology is the Bayesian-specific tension between prior regularization and available signal for weakly-identified hierarchical parameters. In the analyses presented here, that tension was mostly resolved favorably: both real-data applications showed substantial posterior updating for key dispersion and mixing parameters, and the simulation study was generally well-calibrated for ``realistic'' data-generating processes. However, the simulation study revealed that inference for dispersion parameters can become sensitive to choice of prior when the truth lies near the boundary of the parameter space. We therefore view the method as one whose performance depends, like other hierarchical models, on a workable alignment of prior specification, data resolution, and strength of inferential signal.

By contrast, the most immediate limitation for real-world application is the assumption that reported incidence is an adequate stand-in for true incidence. Though this assumption is common in surveillance modeling, it is especially consequential here because the interpretation of $\hat{Y}_{tgi}$ as infectious prevalence, and of $r_{tgi}$ as realized infectiousness, are predicated on perfect reporting. When reporting is incomplete, delayed, or systematically heterogeneous, these latent quantities are better interpreted as proxies for infectious potential, and transmission parameters may be confounded with reporting mechanisms. 

Several directions for future research follow naturally from these and other secondary limitations. Short-term research goals include systematic prior sensitivity analysis for key estimands and improved computation for the latent offspring layer. Long-term research goals include explicit treatment of reporting processes, such as that presented in \citet{2020-Bracher-Held-Underreporting-EE}, and characterization of classes of functions $h_{\bm{γ}}$ that yield identifiable infection-duration and depletion structures; and, (v) accounting for long-run demographic changes including aging, births, deaths, and immigration. More broadly, we hope future research will determine how far the framework can be pushed under finer demographic stratification before the gains in scientific granularity are outweighed by loss of inferential signal.

We view the present work as a step toward surveillance models that are simultaneously stochastic, interpretable, and flexible enough to learn transmission dynamics from fine grain incidence data. The value in this framework is not only that it generalizes the endemic-epidemic model mathematically, but that it also demonstrates the practical viability of fitting richly-parameterized models to doubly-stratified surveillance data. In that sense, our contribution is both methodological and empirical: it shows that doubly-stratified incidence data can support learning about mixing and transmission heterogeneity, provided that regularization, disease-specific assumptions, and latent approximations are made explicit.

%%%%%%%%%%%%%%%%%%%%%%%%%%%%%%%%%%%%%%%%%%%%%%%%%%%%%%%%%%%%%%%%%%%%%%%%%%%%%%%%
%%%%%%%%%%%%%%%%%%%%%%%%%%%%%%%%%%%%%%%%%%%%%%%%%%%%%%%%%%%%%%%%%%%%%%%%%%%%%%%%

\backmatter

\bibliographystyle{biom}
\bibliography{biblio}
%%%%%%%%%%%%%%%%%%%%%%%%%%%%%%%%%%%%%%%%%%%%%%%%%%%%%%%%%%%%%%%%%%%%%%%%%%%%%%%%
%%%%%%%%%%%%%%%%%%%%%%%%%%%%%%%%%%%%%%%%%%%%%%%%%%%%%%%%%%%%%%%%%%%%%%%%%%%%%%%%

\section*{Acknowledgments}

Authors Moran and Trangucci were supported by CDC grant 5U01IP001138-05 via subaward SUBK00022005. The authors thank the staff of the OSU COE high-performance computing cluster for maintaining the computational infrastructure used in this work, and the Michigan Department of Health \& Human Services for providing the COVID-19 incidence data analyzed herein. \vspace*{-8pt}

%%%%%%%%%%%%%%%%%%%%%%%%%%%%%%%%%%%%%%%%%%%%%%%%%%%%%%%%%%%%%%%%%%%%%%%%%%%%%%%%
%%%%%%%%%%%%%%%%%%%%%%%%%%%%%%%%%%%%%%%%%%%%%%%%%%%%%%%%%%%%%%%%%%%%%%%%%%%%%%%%

\section*{Supplementary Materials}

Code for \Cref{s:simstudy,s:inf-noro}, and incomplete code for \Cref{s:inf-covid} is available as a GitHub repository, located at \href{https://github.com/MilesMoran/bayes-epi-models}{\texttt{https://github.com/MilesMoran/bayes-epi-models}}.\vspace*{-8pt}

%%%%%%%%%%%%%%%%%%%%%%%%%%%%%%%%%%%%%%%%%%%%%%%%%%%%%%%%%%%%%%%%%%%%%%%%%%%%%%%%
%%%%%%%%%%%%%%%%%%%%%%%%%%%%%%%%%%%%%%%%%%%%%%%%%%%%%%%%%%%%%%%%%%%%%%%%%%%%%%%%

%%%%%%%%%%%%%%%%%%%%%%%%%%%%%%%%%%%%%%%%%%%%%%%%%%%%%%%%%%%%%%%%%%%%%%%%%%%%%%%%
%%%%%%%%%%%%%%%%%%%%%%%%%%%%%%%%%%%%%%%%%%%%%%%%%%%%%%%%%%%%%%%%%%%%%%%%%%%%%%%%
\label{lastpage}

\newpage
\appendix % don't touch this
\section*{\textbf{APPENDIX}}

\setcounter{secnumdepth}{1}
    %%%%%%%%%%%%%%%%%%%%%%%%%%%%%%%%%%%%%%%%%%%%%%%%%%%%%%%%%%%%%%%%%%%%%%%%%%%%%%%%
%%%%%%%%%%%%%%%%%%%%%%%%%%%%%%%%%%%%%%%%%%%%%%%%%%%%%%%%%%%%%%%%%%%%%%%%%%%%%%%%

\section{Beta-Binomial Parameterization}
\label[webappendix]{WA:beta-binom}

Here, we clarify the notation used in \Cref{ss:model-def-covid} to refer to the Beta-Binomial distribution. Let $N$ denote the binomial sample size, $p∈(0,1)$ the mean success probability, and $k∈(0,∞)$ the precision. We use $X \sim \mathrm{BetaBinom}(N,p,k)$ to refer to a random variable with support $\{0,...,N\}$ and probability mass function 
\begin{equation*}
    \Pr{X=x|N,p,k} = \binom{N}{x} \frac{B(x+pk, N-x+(1-p)k)}{B(pk, (1-p)k)}
\end{equation*}
where $B(\cdot,\cdot)$ denotes the standard Beta function. This distribution has mean $\E{}{X}=Np$ and variance $\Var{X} = Np(1-p)\left(\frac{k+N}{k+1}\right)$, thus representing an overdispersed binomial distribution for small $k$ and nesting the exact $\Bin{N,p}$ distribution when $k\uparrow ∞$.

%%%%%%%%%%%%%%%%%%%%%%%%%%%%%%%%%%%%%%%%%%%%%%%%%%%%%%%%%%%%%%%%%%%%%%%%%%%%%%%%
%%%%%%%%%%%%%%%%%%%%%%%%%%%%%%%%%%%%%%%%%%%%%%%%%%%%%%%%%%%%%%%%%%%%%%%%%%%%%%%%

    %%%%%%%%%%%%%%%%%%%%%%%%%%%%%%%%%%%%%%%%%%%%%%%%%%%%%%%%%%%%%%%%%%%%%%%%%%%%%%%%
%%%%%%%%%%%%%%%%%%%%%%%%%%%%%%%%%%%%%%%%%%%%%%%%%%%%%%%%%%%%%%%%%%%%%%%%%%%%%%%%

\section{Hyperparameters for Priors of Age-Group Mixing Parameters}
\label[webappendix]{WA:clarify-wstrata-priors}

In \Cref{ss:model-wstrata-prior}, we describe how the generative prior scheme for age-group mixing parameters induces elementwise gamma priors on the contact rate matrix $\bm{C}$. We detail that $α_{ii}>α_{ii'}$ is used to ``loosely-regularize'' towards assortative mixing for the model fits presented herein. Specifically, assume $α_1$ is the shape parameter for diagonal elements and $α_2<α_1$ is that for off-diagonal elements. Leveraging the induced Dirichlet prior, the diagonal elements of $\bm{w}^{(I)}$ have mean
\begin{align*}
    \E{}{w^{(I)}_{ii}} = \frac{α_1}{α_1 + (I-1)α_2} 
        = \frac{\mathrm{conc}_{\mathrm{diag}}}
               {\mathrm{conc}_{\mathrm{total}}}.
\end{align*}
From this, we can parameterize the elements of $\bm{w}^{(I)}$ in terms of the expected diagonal proportion $\E{}{w^{(I)}_{ii}}$ and the diagonal concentration. For our inference, we set $\E{}{w^{(I)}_{ii}}=0.4$ and $α_{1}=\mathrm{conc}_{\mathrm{diag}}=1.8$ consistent with the contact matrix used in \Cref{s:simstudy}. With $I=6$, this leads to shape parameters $α_{1}≈4.32$ and \mbox{$α_{2}≈1.30$}, which regularizes towards assortative mixing structures.

%To address posterior curvature and ensure stable HMC sampling, we refuse direct sampling in favor of a prior scheme that intentionally decouples the composition of $\bm{C}$ (i.e. $\bm{w}^{(I)}$) from the nuisance column scales. In this parameterization, the $i^{\mathrm{th}}$ column of the lower-triangle of $\bm{C}$ is written sequentially as the product of a scalar parameter $u_{i}>0$ and a simplex $\bm{q}_{i}$ of size $(I-i+1)$. By choosing specific gamma priors for $\{u_{i}\}$ and dirichlet priors for $\{\bm{q}_{i}\}$, we induce the desired independent elementwise priors \mbox{$C_{ii'} \sim \mathrm{Gamma}(α_{ii'},\tfrac{1}{2I})$}. 

%%%%%%%%%%%%%%%%%%%%%%%%%%%%%%%%%%%%%%%%%%%%%%%%%%%%%%%%%%%%%%%%%%%%%%%%%%%%%%%%
%%%%%%%%%%%%%%%%%%%%%%%%%%%%%%%%%%%%%%%%%%%%%%%%%%%%%%%%%%%%%%%%%%%%%%%%%%%%%%%%

    %%%%%%%%%%%%%%%%%%%%%%%%%%%%%%%%%%%%%%%%%%%%%%%%%%%%%%%%%%%%%%%%%%%%%%%%%%%%%%%%
%%%%%%%%%%%%%%%%%%%%%%%%%%%%%%%%%%%%%%%%%%%%%%%%%%%%%%%%%%%%%%%%%%%%%%%%%%%%%%%%

\section{Simulation Study Log-Score Prediction Evaluation}
\label[webappendix]{WA:clarify-log-score}

Recall that $d∈\{1,...,D\}$-many datasets were generated for each DGP scenario $s∈\{1,...,S\}$ in the simulation study (where $D=200$ and $s=16$). We now formalize the model evaluation process for a single scenario $s$. For each $d$, let $\bm{Y}^{*(d)}_{1:T}$ denote the \mbox{$T×G×I$} array of counts forming the training dataset. Let $\bm{Y}^{*(d)}_{T+h}$ denote the \mbox{$G×I$} matrix of counts forming the $h$-horizon forecasting target, for each $h∈\{1,...,H\}$. For posterior draw $k∈\{1,...,K\}$, conditional independence implies a factorization of the $h$-step-ahead posterior-predictive density
\begin{equation*}
    p\Big(\bm{Y}^{*(d)}_{T+h} \,\Big|\, \bm{Θ}^{(d,k)}, \bm{Y}^{*(d)}_{1:T} \Big) 
        = \prod_{g,i} p\Big(
            Y^{*(d)}_{(T+h)gi} \,\Big|\, \bm{Θ}^{(d,k)}, \bm{Y}^{*(d)}_{1:T}
        \Big),
\end{equation*}
with univariate products on the right side given by the model's likelihood (negative binomial). Let $\mathrm{LS}^{(d)}_{m,h}$ denote the log-score of model $m$'s $h$-step-ahead predictive density for dataset $d$. We use the factorization above to estimate log-score via Monte Carlo integration:
\begin{equation*}
    \widehat{\mathrm{LS}}^{(d)}_{m,h}
        = \log{ \frac{1}{K} \sum_{k=1}^{K}
                p_m \Big( 
                    \bm{Y}^{*(d)}_{T+h} \,\Big|\, \bm{Θ}_{m}^{(d,k)}, \bm{Y}^{*(d)}_{1:T}
                \Big)
            }.
\end{equation*}
We use $Δ_{h}^{(d)} = \widehat{\mathrm{LS}}^{(d)}_{\mathrm{full},h} - \widehat{\mathrm{LS}}^{(d)}_{\mathrm{reduced},h}$ as the per-dataset relative-performance of the reduced model, so that $Δ_{h}^{(d)}>0$ indicates predictive superiority of the proposed model over the reduced model at horizon $h$ for dataset $d$. We report the mean over datasets $\bar{Δ}_{h}$ and its standard error:
\begin{align*}
    \mathrm{SE}( \bar{Δ}_{h} ) 
        &= \sqrt{\frac{1}{D(D-1)}\sum_{d=1}^{D}(Δ_{h}^{(d)} - \bar{Δ}_{h})^2}.
\end{align*}
This last step is repeated for the naive model, and the entire procedure is repeated for each DGP scenario. Note that our reported variability does not include error in the monte carlo estimate of the log-score, as this added variability is negligible relative to between-datasets variability.

%%%%%%%%%%%%%%%%%%%%%%%%%%%%%%%%%%%%%%%%%%%%%%%%%%%%%%%%%%%%%%%%%%%%%%%%%%%%%%%%
%%%%%%%%%%%%%%%%%%%%%%%%%%%%%%%%%%%%%%%%%%%%%%%%%%%%%%%%%%%%%%%%%%%%%%%%%%%%%%%%

    %%%%%%%%%%%%%%%%%%%%%%%%%%%%%%%%%%%%%%%%%%%%%%%%%%%%%%%%%%%%%%%%%%%%%%%%%%%%%%%%
%%%%%%%%%%%%%%%%%%%%%%%%%%%%%%%%%%%%%%%%%%%%%%%%%%%%%%%%%%%%%%%%%%%%%%%%%%%%%%%%

\section{Equivalent Mean Structures in Rare Disease Context}
\label[webappendix]{WA:proof-means}

Consider the EE model discussed in \Cref{ss:model-def-noro} alongside its immediate extension where $θ>0$ (still retaining all other rare-disease assumptions). The introduction of latent transmission heterogeneity does not impact marginal mean incidence as $λ_{tgi}$ is linear in $\bm{r}_{t}$. 
\begin{proof}
By the law of total expectation, we have
\begin{align*}
    \E{θ>0}{Y^{*}_{tgi} \mid \bm{Y}^{*}_{t-1}}
        &= \E{}{\E{}{Y^{*}_{tgi} \mid \bm{r}_{t}} \mid \bm{Y}^{*}_{t-1}} \\
        &= \E{}{λ_{tgi}(\bm{r}_{t}) \mid \bm{Y}^{*}_{t-1}} \\
        &= λ_{tgi}\left(\E{}{\bm{r}_t \mid \bm{Y}^{*}_{t-1}}\right) \\
        &= λ_{tgi}\left(R_0\bm{Y}^{*}_{t-1}\right) \\
        &= \E{θ=0}{Y^{*}_{tgi} \mid \bm{Y}^{*}_{t-1}}.
\end{align*}    
\end{proof}

%%%%%%%%%%%%%%%%%%%%%%%%%%%%%%%%%%%%%%%%%%%%%%%%%%%%%%%%%%%%%%%%%%%%%%%%%%%%%%%%
%%%%%%%%%%%%%%%%%%%%%%%%%%%%%%%%%%%%%%%%%%%%%%%%%%%%%%%%%%%%%%%%%%%%%%%%%%%%%%%%

    %%%%%%%%%%%%%%%%%%%%%%%%%%%%%%%%%%%%%%%%%%%%%%%%%%%%%%%%%%%%%%%%%%%%%%%%%%%%%%%%
%%%%%%%%%%%%%%%%%%%%%%%%%%%%%%%%%%%%%%%%%%%%%%%%%%%%%%%%%%%%%%%%%%%%%%%%%%%%%%%%

\section{Inter-PUMA Distance Matrix}
\label[webappendix]{WA:puma-distmat}

In \Cref{s:inf-covid}, we introduce a new ``distance'' quantity $D_{gg'}$ used to define between-geographies mixing structure in-lieu of the adjacency-order construction presented in \cite{2017-Meyer-Held-Social-Contact}. Here, we elaborate that the distance $D_{gg'}$ between PUMAs $g$ and $g'$ is calculated as the average Haversine distance between two randomly-chosen individuals -- one from PUMA $g$ and one from PUMA $g'$ -- in tens of kilometers. Because individual location is aggregated up to the level of census tracts (a fine-grain partition of the PUMAs), this ``human distance'' is an approximation made by assuming individuals are all located at their respective census tracts' centroids. 

Formally, let $E_{g}$ denote the population count of PUMA $g$ and $e_{u}$ the population count of census tract $u$. Noting that census tracts form a partition of a PUMA, let $\text{Tracts}(g)$ denote the set of census tracts that comprise PUMA $g$. Let $d_{uv}$ denote the Haversine distance (in km) between the centroids of tracts $u$ and $v$. We define $D_{gg'}$ by the formula
\begin{equation*}
    D_{gg'} 
        = \frac{1}{10} 
            \sum_{\substack{u∈\text{Tracts}(g)\\ v∈\text{Tracts}(g')}}
            \frac{e_{u}e_{v}}{E_{g}E_{g'}} d_{uv},
\end{equation*}
where $e_{u}/E_{g}$ denotes the probability a random sample of persons in PUMA $g$ yields someone from census tract $u$ (and similar for $e_{v}/E_{g'}$).

This notion of human distance between PUMAs is straightforward, albeit incomplete. Because census tracts form a fine partition of the study region ($≈3000$ tracts vs $66$ PUMAs across the MI lower-peninsula), the definition above controls for variation in PUMA shape, size, and spatial concentration when compared to the naive approach of using distance between PUMA centroids. Still, a better alternative for constructing distance deterministically would involve extracting $D_{gg'}$ directly from existing commuting-flow data, e.g. from the U.S. Census Bureau’s LEHD Origin-Destination Employment Statistics (``LODES'').

%%%%%%%%%%%%%%%%%%%%%%%%%%%%%%%%%%%%%%%%%%%%%%%%%%%%%%%%%%%%%%%%%%%%%%%%%%%%%%%%
%%%%%%%%%%%%%%%%%%%%%%%%%%%%%%%%%%%%%%%%%%%%%%%%%%%%%%%%%%%%%%%%%%%%%%%%%%%%%%%%

    \newpage
    %%%%%%%%%%%%%%%%%%%%%%%%%%%%%%%%%%%%%%%%%%%%%%%%%%%%%%%%%%%%%%%%%%%%%%%%%%%%%%%%
%%%%%%%%%%%%%%%%%%%%%%%%%%%%%%%%%%%%%%%%%%%%%%%%%%%%%%%%%%%%%%%%%%%%%%%%%%%%%%%%

\section{Priors for Simulation Study \& Berlin Norovirus Inference}
\label[webappendix]{WA:clarify-other-priors}

\begin{table}[!ht]
\addtolength{\tabcolsep}{-1em}
\linespread{1.3}\selectfont
\centering
\rowcolors{2}{white}{mylightgray}

\resizebox{\linewidth}{!}{ $\displaystyle
\begin{array}{l>{\collectcell\texttt}l<{\endcollectcell}cccll} \hline
    \textbf{Param.} & \textbf{\textrm{Stan Name}} & \textbf{N} & \textbf{R} & 
        \textbf{F} & \textbf{Prior} & \textbf{Description} \\ \hline
    β_0                 & end\_1            & × & × & × & \Normal{3,2}      &
        \text{endemic log-rate of reference-group $(g,i)=(1,1)$, seasonal avg.} \\
    β_{g}^{(G)}         & end\_geogs[g]     & × & × & × & \Normal{0,3}      &
        \text{endemic log-rate geog. $g$ additive shift (rel. to $g=1$)}    \\
    β_{i}^{(I)}         & end\_strata[i]    & × & × & × & \Normal{0,3}      &
        \text{endemic log-rate strata $i$ additive shift (rel. to $i=1$)}   \\
    β_{i}^{(S)}         & end\_sin[i]       & × & × & × & \Normal{0,3}      &
        \text{endemic log-rate strata-specific sine-seasonality}            \\
    β_{i}^{(C)}         & end\_cos[i]       & × & × & × & \Normal{0,3}      &
        \text{endemic log-rate strata-specific cosine-seasonality}          \\        
    β^{(\mathrm{xmas})} & end\_christmas    & × & × & × & \Normal{0,3}      &
        \text{endemic log-rate extra holiday ``shock'' additive shift}      \\
    -\log{ψ}            & nlog\_overdisp    & × & × & × & \Normal{-\tfrac{1}{2},1} &
        \text{(nlog of) overdispersion of observation-level neg. bin. distribution} \\
    η_{0}               & ne\_1             &   & × & × & \Normal{2,5}      &
        \text{log-susceptibility of reference-group $(g,i)=(1,1)$}          \\
    η_{g}^{(G)}         & ne\_geogs[g]      &   & × & × & \Normal{0,3}      &
        \text{log-susceptibility geog. $g$ additive shift (rel. to $g=1$)}  \\
    η_{i}^{(I)}         & ne\_strata[i]     &   & × & × & \Normal{0,3}      &
        \text{log-susceptibility strata $i$ additive shift (rel. to $i=1$)} \\
    η^{(\mathrm{pop})}  & ne\_log\_pop      &   & × & × & \Normal{0,2}      &
        \text{log-susceptibility population-share coefficient}              \\
    \log{ρ}             & neweights\_logd   &   & × & × & \Normal{0,1}      &
        \text{(log of) global spatial distance (adjacency-order) decay rate} \\
    \log{κ}             & logpower          &   & × &   & \Normal{0,\tfrac{3}{4}} &
        \text{(log of) Meyer-Held contact matrix eigenvalues power}         \\
    -\log{θ}            & nlog\_theta       &   &   & × & \Normal{-2,1}     &
        \text{(nlog of) scale of latent offspring distribution}             \\ \hline
\end{array}
$}

\caption{Priors for parameters seen in the simulation study. Parameters are listed alongside the corresponding ASCII name used in Stan. Presence of a parameter in the naive (N), reduced (R), or full (F) model is denoted by $×$. Priors for mixing weights are introduced in \Cref{ss:model-wstrata-prior} and described in greater detail in \Cref{WA:clarify-wstrata-priors}. Note that the priors for $θ$ and $ψ$ differ between simulation study and norovirus dataset analysis (see \Cref{ss:noro-data-description})} 
\label{tbl:other-priors}
\end{table}

%%%%%%%%%%%%%%%%%%%%%%%%%%%%%%%%%%%%%%%%%%%%%%%%%%%%%%%%%%%%%%%%%%%%%%%%%%%%%%%%
%%%%%%%%%%%%%%%%%%%%%%%%%%%%%%%%%%%%%%%%%%%%%%%%%%%%%%%%%%%%%%%%%%%%%%%%%%%%%%%%

    %%%%%%%%%%%%%%%%%%%%%%%%%%%%%%%%%%%%%%%%%%%%%%%%%%%%%%%%%%%%%%%%%%%%%%%%%%%%%%%%
%%%%%%%%%%%%%%%%%%%%%%%%%%%%%%%%%%%%%%%%%%%%%%%%%%%%%%%%%%%%%%%%%%%%%%%%%%%%%%%%

\section{Priors for Michigan COVID-19 Inference}
\label[webappendix]{WA:clarify-covid-priors}

\begin{table}[!ht]
\addtolength{\tabcolsep}{1em}
\linespread{1.3}\selectfont
\centering
\rowcolors{2}{white}{mylightgray}

\resizebox{\linewidth}{!}{ $\displaystyle
\begin{array}{l>{\collectcell\texttt}l<{\endcollectcell}ll} \hline
    \textbf{Param.} & \textbf{\textrm{Stan Name}} & \textbf{Prior}          & 
        \textbf{Description}                                                \\ \hline
    δ_{gi}              & delta[g,i]        & \mathrm{HalfNormal}(σ=1)      &
        \text{endemic FOI background rate}                                  \\
    θ                   & theta             & \mathrm{HalfCauchy}(σ=5)      &
        \text{scale of latent offspring distribution}                       \\ 
    k                   & k                 & \mathrm{HalfCauchy}(σ=10^{4}) &
        \text{precision of observation-level beta-binomial distribution}    \\
    γ                   & gamma             & \mathrm{HalfCauchy}(σ=1)      &
        \text{recovery rate in disease prevalence estimate}                 \\
    \log{R_{t}}         & log\_R[t]         & \mathrm{Normal}(0,2)          &
        \text{(log of) latent offspring mean multiplier (in time)}          \\
    \log{α_{i}}         & log\_alpha[i]     & \mathrm{Gamma}(10.8,12)       &
        \text{(log of) latent offspring mean multiplier (in age strata)}    \\    
    \log{τ_{g}}         & log\_tau[g]       & \mathrm{Normal}(0,3)          &
        \text{(log of) destination-specific spatial attractiveness}         \\
    μ_{\mathrm{log}\,ρ} & mean\_log\_rho    & \mathrm{Normal}(0,2)          &
        \text{hierarchical mean for random effects } \log{ρ_{g}}            \\
    σ_{\mathrm{log}\,ρ} & sd\_log\_rho      & \mathrm{HalfNormal}(σ=1)      &
        \text{hierarchical std. dev. for random effects } \log{ρ_{g}}       \\
    \log{ρ_{g}}         & rho[g]            & \mathrm{Normal}(μ_{\mathrm{log}ρ},
                                                      σ_{\mathrm{log}ρ})    &
        \text{(log of) source-specific spatial distance decay rate}         \\ \hline
\end{array}
$}

\caption{Priors for parameters seen in the outbreak-driven model specification of \Cref{ss:model-def-covid}. Parameters are listed alongside the corresponding ASCII name used in Stan. Priors for mixing weights are introduced in \Cref{ss:model-wstrata-prior} and described in greater detail in \Cref{WA:clarify-wstrata-priors}.}  
\label{tbl:covid-priors}
\end{table}

%%%%%%%%%%%%%%%%%%%%%%%%%%%%%%%%%%%%%%%%%%%%%%%%%%%%%%%%%%%%%%%%%%%%%%%%%%%%%%%%
%%%%%%%%%%%%%%%%%%%%%%%%%%%%%%%%%%%%%%%%%%%%%%%%%%%%%%%%%%%%%%%%%%%%%%%%%%%%%%%%

\clearpage
\onecolumn
\begin{figure*}[!t]
    \centering
    \linespread{1.3}\selectfont
    \includegraphics[width=\textwidth]{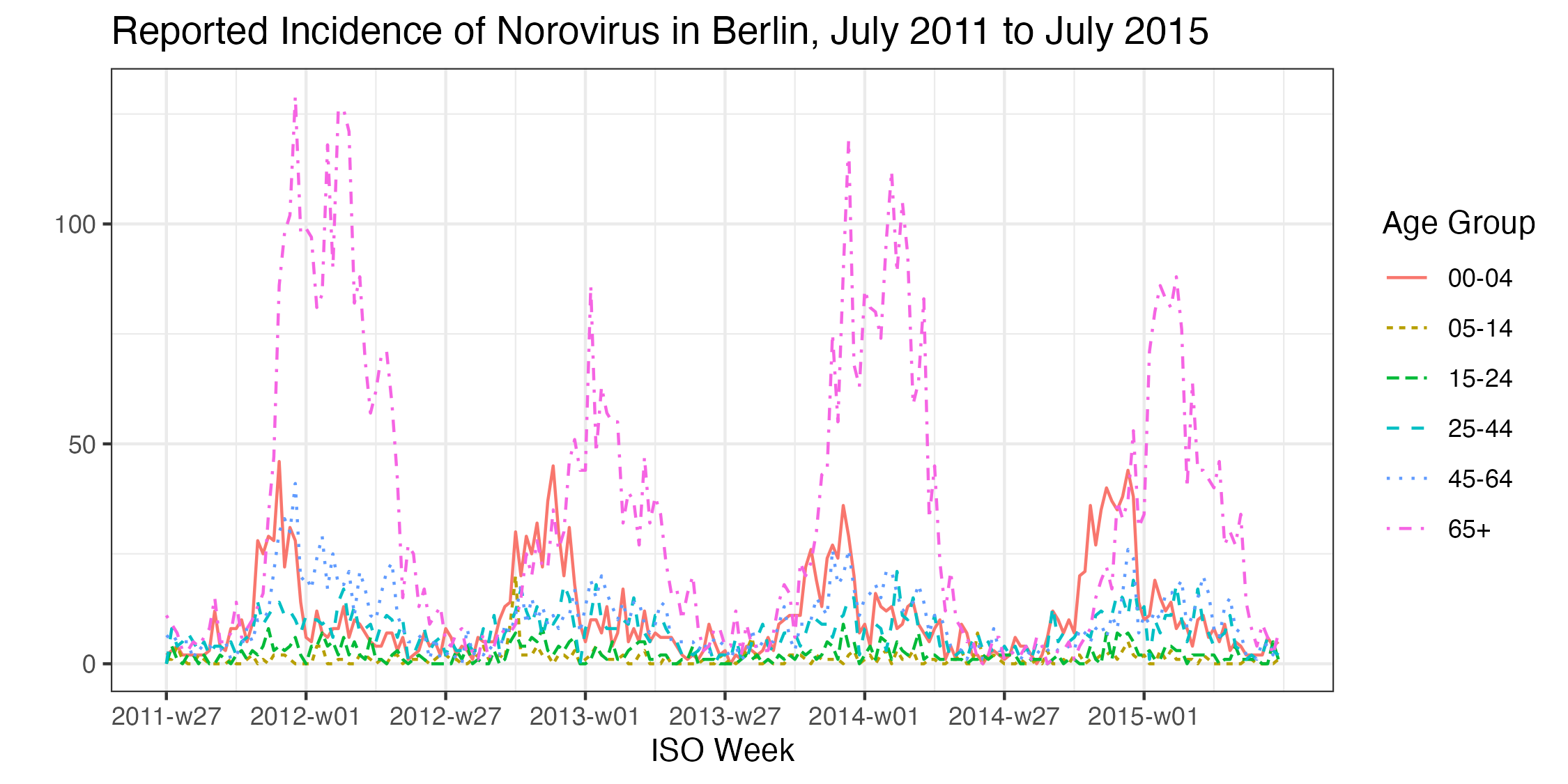}
    \includegraphics[width=\textwidth]{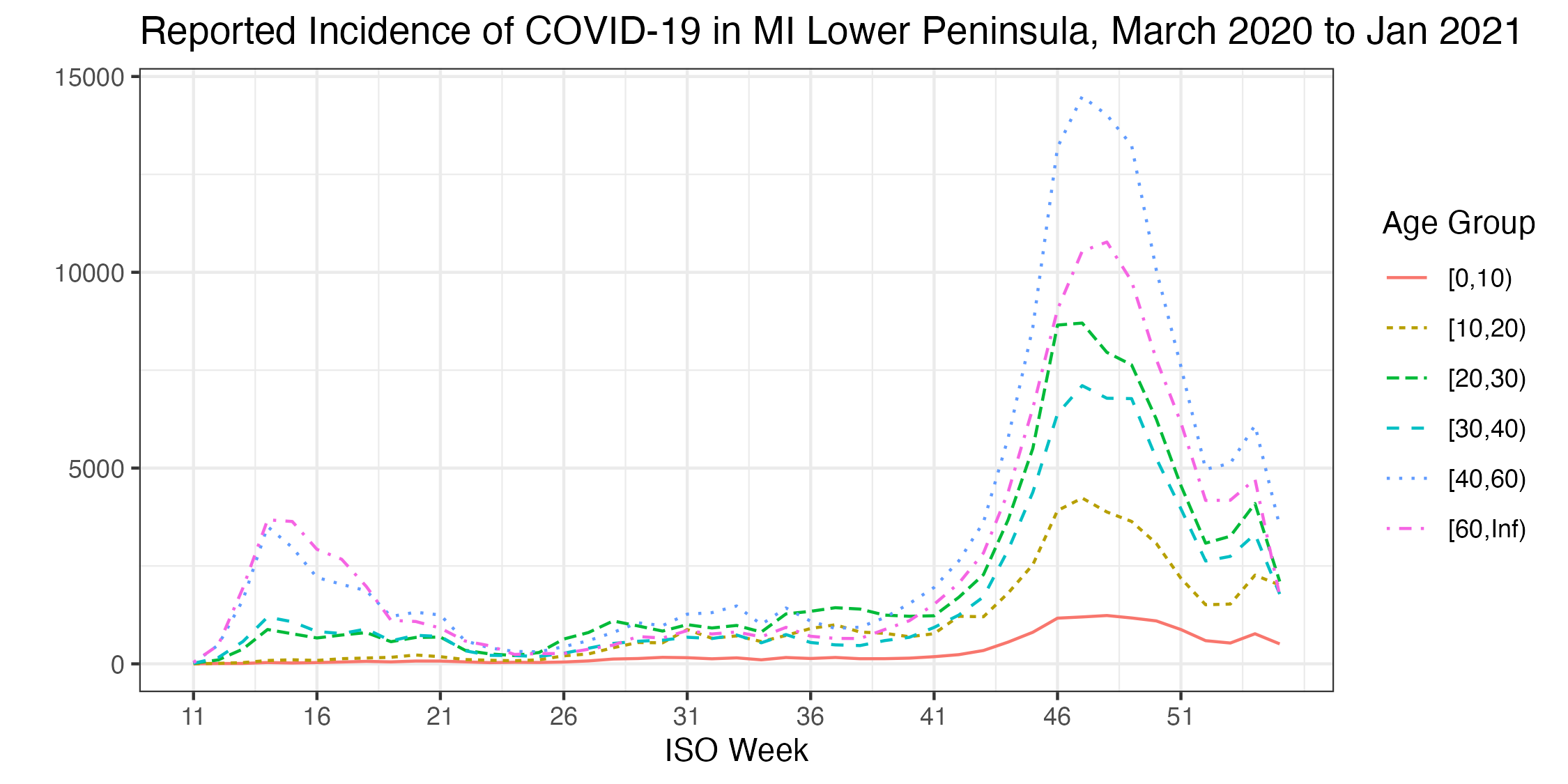}
    \caption{Incident disease counts for both analyses, aggregated over geographic region. This comparison is intended to highlight the unique ``outbreak-driven'' behavior of the Michigan COVID-19 dataset and motivate the need for model features distinct from those of the highly-endemic Berlin Norovirus dataset. Importantly, the gradual decrease of COVID-19 incidence after week 13 coincides \textit{exactly} with the introduction of stay-at-home orders by local authorities; however, the drastic rise in incidence after week 41 comes long after these orders were relieved in July 2020.}
    \label{plot:incidence-comparison}
\end{figure*}

\newpage
\begin{figure*}[!t]
    \centering
    \linespread{1.3}\selectfont
    \includegraphics[width=\textwidth]{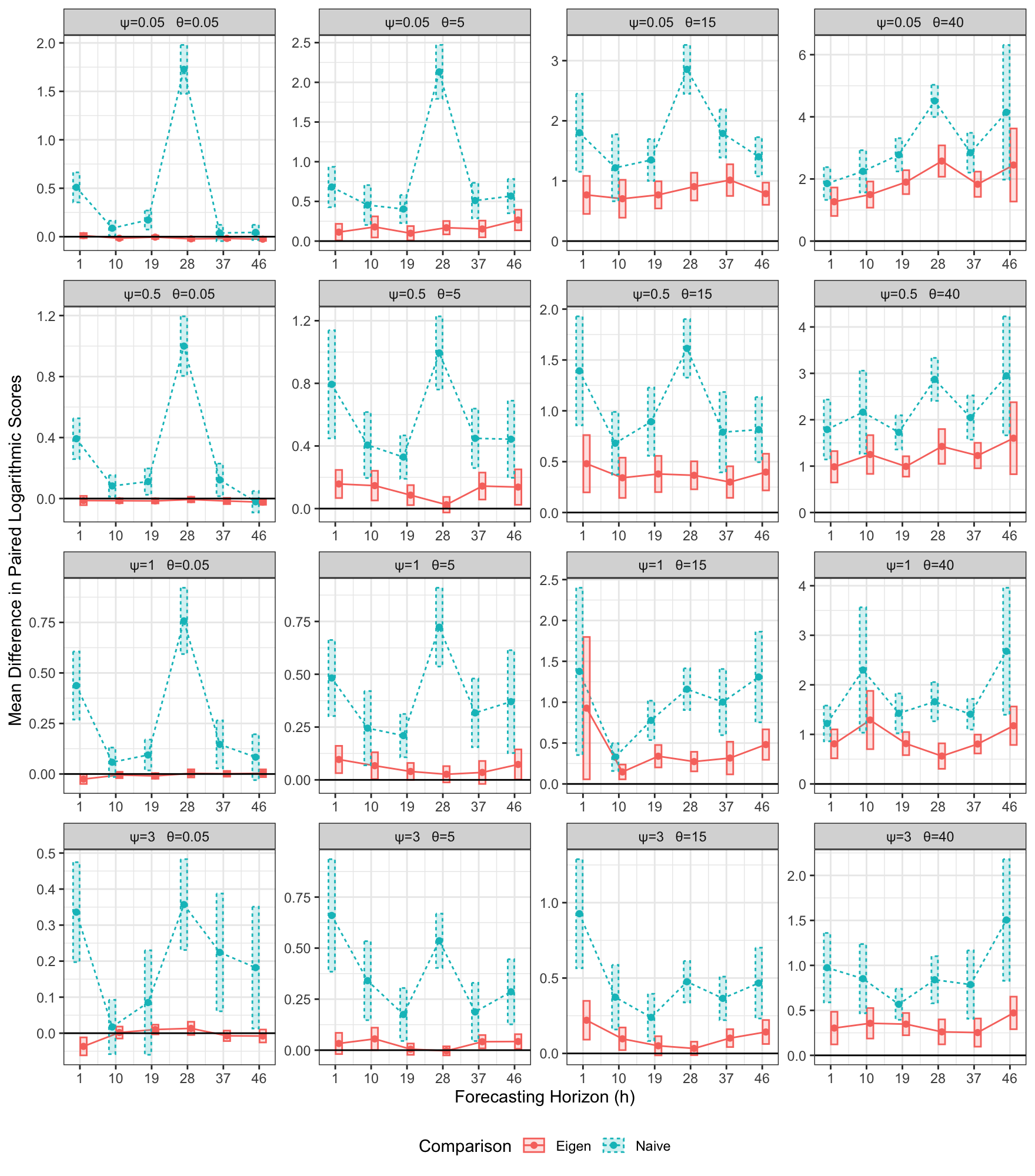}
    \caption{Mean difference in paired log-scores. Intervals for the full-vs-naive comparison are given in blue (dashed lines), and intervals for the full-vs-reduced comparison are given in red (solid lines). Computation of these intervals is described in \Cref{WA:clarify-log-score}. Intervals are offset (``jittered'') along the $x$-axis for visual separability.}
    \label{plot:simstudy-logscores-full}
\end{figure*}

\newpage
\begin{figure*}[!t]
    \linespread{1.3}\selectfont
    \hspace{-9mm}
    \includegraphics[width=1.05\linewidth]{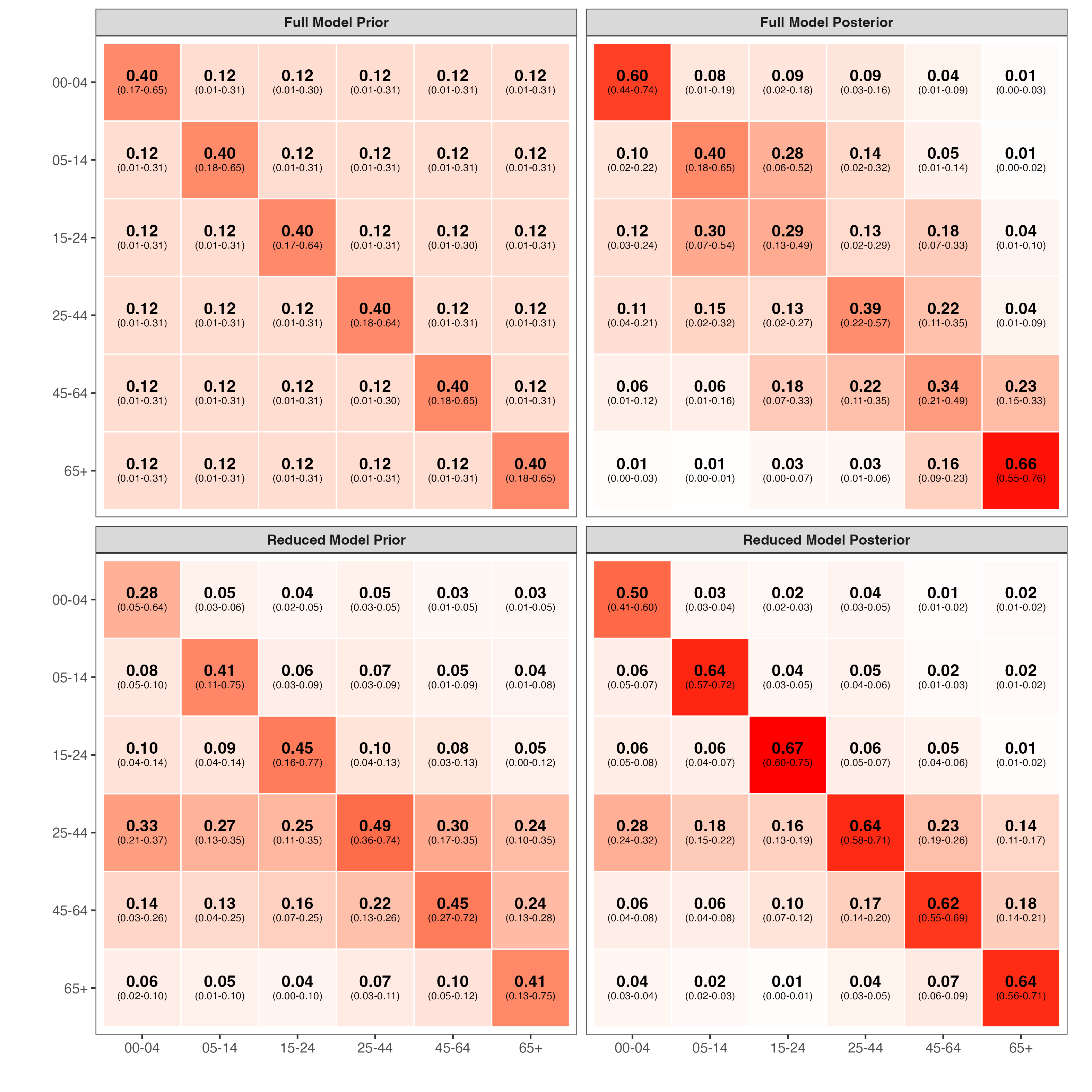}
    \caption{Prior and posterior means (and 90\% equal-tailed intervals) for elements of $\bm{w}^{(I)}$ from the Norovirus model fit. Origin age-groups are depicted on the x-axis and recipients on the y-axis. All matrices are column-stochastic.}    
    \label{plot:inf-noro-wstrata-estm}
\end{figure*}

\end{document}